\documentclass[prl,aps,twocolumn,superscriptaddress,preprintnumbers,showpacs,floatfix,amsmath,amssymb]{revtex4}

\usepackage{graphicx}% Include figure files
\usepackage{dcolumn}% Align table columns on decimal point
\usepackage{bm}% bold math

\usepackage{amsmath}
\usepackage{graphicx}
\usepackage{amssymb}
\usepackage{mathrsfs}
\usepackage{bbm}
\usepackage{epsfig}
\newcommand{\la}{\label}

\newcommand{\be}{\begin{eqnarray}}
\newcommand{\ee}{\end{eqnarray}}

\begin{document}

%\pagenumbering{empty}
%\begin{titlepage}
%
\title{  Bloch spin waves and emergent structure in protein folding \\ 
with HIV envelope glycoprotein as an example 
}

%\vskip 5.0cm
%\vskip 5.0cm
\author{Jin Dai} 
\email{daijing491@gmail.com}
\affiliation{School of Physics, Beijing Institute of Technology, Beijing 100081, P.R. China}
\author{Antti J. Niemi}
\email{Antti.Niemi@physics.uu.se}
\affiliation{Department of Physics and Astronomy, Uppsala University,
P.O. Box 803, S-75108, Uppsala, Sweden}
\affiliation{
Laboratoire de Mathematiques et Physique Theorique
CNRS UMR 6083, F\'ed\'eration Denis Poisson, Universit\'e de Tours,
Parc de Grandmont, F37200, Tours, France}
\affiliation{School of Physics, Beijing Institute of Technology, Beijing 100081, P.R. China}
\homepage{http://www.folding-protein.org}
\author{Jianfeng He}
\email{hjf@bit.edu.cn}
\affiliation{School of Physics, Beijing Institute of Technology, Beijing 100081, P.R. China}
\author{Adam Sieradzan}
\email{adams86@wp.pl}
\affiliation{Faculty of Chemistry, University of Gdansk, Wita Stwosza 63, 80-308 Gda\'nsk, Poland}
\author{Nevena Ilieva}
\email{nilieval@mail.cern.ch}
\affiliation{Institute of Information and Communication Technologies, 25A, Acad. G. Bonchev Str., Sofia 1113, Bulgaria}

\begin{abstract}
\noindent
We inquire how structure emerges during the process of protein folding. 
For this we scrutinise  collective many-atom motions during
all-atom molecular dynamics simulations. We introduce, develop and employ
various topological  techniques, in combination with analytic tools that we deduce from 
the concept of integrable models and structure of discrete nonlinear Schr\"odinger equation. 
The example we consider is an $\alpha$-helical subunit of the HIV envelope glycoprotein gp41. 
The helical structure is stable when the subunit is part of the biological
oligomer. But in isolation the  helix becomes unstable, and the monomer starts deforming. 
We follow the process computationally. We interpret the evolving structure both in terms 
of a backbone based Heisenberg spin chain and in terms of a side chain based XY spin chain.
We find that in both cases the formation of protein super-secondary structure is akin the formation of a topological 
Bloch domain wall along a spin chain. During the 
process we identify three individual Bloch walls and we show that each of them can be modelled with a very high 
precision in terms of  a soliton solution to a discrete nonlinear 
Schr\"odinger  equation. 
\end{abstract}


%\date{\today}

\maketitle

\section{Introduction}

A domain wall is a prototype collective excitation in a physical system,
and it is also the paradigm example of a topological soliton \cite{Manton-2004}. 
A domain wall can appear  whenever there is a global symmetry that 
becomes spontaneously broken.
It constitutes the boundary that separates two 
neighbouring domains, in which the  order parameter that detects the 
symmetry breaking has different values. 

In the case of a one dimensional Heisenberg spin chain the order 
parameter is a three component unit length vector. When one of the three vector components vanishes identically,
the Heisenberg spin chain reduces to the XY spin chain \cite{Faddeev-1987,Ablowitz-2004}. 
A domain wall along  the spin chain is  a localised 
excitation that interpolates between two different, ordered spin states in which the 
order parameter has different constant values.  
Two major types of domain walls are commonly identified along the 
Heisenberg chain \cite{Faddeev-1987,Ablowitz-2004}. These are called the Bloch wall and the N\'eel wall, respectively. 
In the case of a Bloch wall, the Heisenberg spin variable rotates through the plane of the wall and
in the case of a N\'eel wall the rotation takes place within the plane of the wall itself. 
Domain walls that are  mixtures of these two, can also occur along the Heisenberg spin 
chain, while along the XY spin chain, only domain walls of the Bloch type can be present.

In this article we demonstrate that the formation of super-secondary structures, during 
folding of a protein \cite{Dill-2012}, can be understood in
terms of a Bloch domain wall that forms along a Heisenberg spin chain, or
along a closely related XY spin chain. 
We propose that the spin chain interpretation of a  protein backbone provides  
a systematic framework for understanding and describing the process of protein  folding. 
For this we employ all-atom force fields \cite{Leach-2001,Frenkel-2001} 
to scrutinise protein folding dynamics at the level of individual atoms and their oscillations. 
We analyse the folding pathway using a combination of topological techniques and global 
analytic tools. We isolate the collective oscillations which are pertinent for the folding process, 
from the noisy background of thermal and random individual atom fluctuations. 
In particular, we illustrate how the individual atom motions become organised and  
combined into a coherent structural excitation which we identify as
the Bloch wall. 

%To construct the local O(3) Heisenberg spin variable, we utilise the C$\alpha$ backbone 
%geometry: The latitude and longitude coordinates of the O(3) spin 
%variable are simply the bond and the torsion angles along the C$\alpha$ backbone.  
%The domain wall is then a loop structure along the protein backbone,  on which the bond
%and torsion angles are variable. The loop interpolates between two adjacent 
%ordered states, that are the regular secondary structures such as $\alpha$ helices,  $\beta$ 
%strands and 3/10 helices. Indeed,  along such 
%regular structures, the bond and torsion angles have essentially 
%constant values.  

As a concrete example we consider an $\alpha$-helical subunit of the HIV envelope 
glycoprotein gp41 \cite{Chan-1997}, with Protein Data Bank \cite{Berman-2000} (PDB) code 
1AIK. There are six $\alpha$-helical subunits in the biological assembly, shown in Figure 1.
We consider in isolation the subunit, for which the first amino acid has number 628 in the PDB file. 
In isolation, the subunit is unstable and starts folding.
%
%
%
%          figure-1b
%
%
%
\begin{figure}
\centering
{\begin{minipage}[t]{.45\textwidth}
  %\raggedright
  \includegraphics[%trim = 0mm 5mm 0mm 90mm, 
  width=0.4\textwidth,  angle=0]{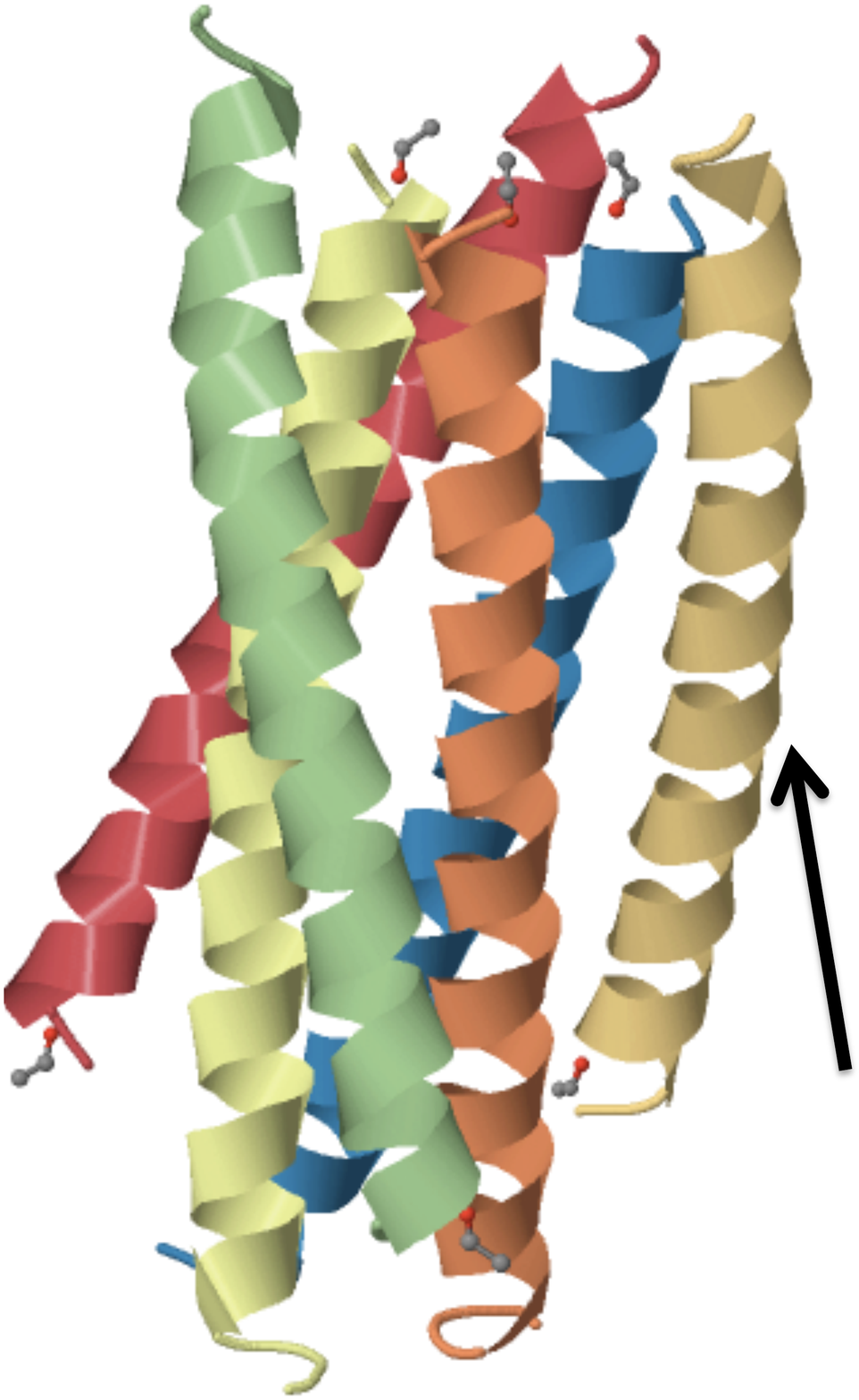}
  \caption{The biological assembly of 1AIK is an oligomer with six $\alpha$-helical structures.
  The subunit that starts with amino acid number 628 in the PDB file is identified by the arrow.}
  \la{fig-1}
\end{minipage}}
\end{figure}

The transmembrane glycoprotein 41 is itself a subunit of the retrovirus envelope 
protein complex.  In the case of the HIV, its structure has been studied extensively. 
It is presumed to have substantial biological relevance to the initial viral infection. 
Accordingly,  the gp41 protein is a popular target for the 
development of  an anti-viral immune response, to prevent and cure
HIV infection. However, medical applications are beyond the direct scope of the present study. 
Here we shall solely address and identify the physical mechanism, why and how an individual,
initially $\alpha$-helical subunit of 1AIK becomes unstable in isolation, and starts folding.

For our all-atom molecular dynamics simulations,  we utilise 
the GROMACS 4.6.3 package \cite{Hess-2008}.  We analyse the results
using  a variety of topological techniques and analytical tools. Our approach 
derives from the mathematical structure  of 
Heisenberg and XY spin chains, in combination with properties of
a discrete nonlinear Schr\"odinger (DNLS) Hamiltonian \cite{Faddeev-1987,Ablowitz-2004}. In particular, the DNLS equation
that describes the local extrema of the Hamiltonian, enables us to analytically identify the profile of
the domain wall, and to interpret it in terms of DNLS soliton \cite{Molkenthin-2011}.

Here we present results from the detailed investigation of a particular example. However, we expect our
observations and conclusions to be generic.  Indeed, the present results are fully in line with the previous findings 
\cite{Krokhotin-2012-a,Krokhotin-2014-a,Sieradzan-2014} 
obtained by using the coarse grained UNRES  energy  function 
\cite{Khalili-2005-a,Khalili-2005-b,Liwo-2005} 
in the case of protein A.
The similitude of results that are obtained by analysing the protein folding process using different tools,
built and based on phenomena with very different characteristic time and length scales,
demonstrates that we have correctly identified the relevant collective motions that
command the folding process.

\section{Methods}

We have performed {\it in silico} experiments to fold one C-chain 
subunit of the core structure of gp41 \cite{Chan-1997}. The structure comes 
from the HIV envelope glycoprotein with PDB code 1AIK. The amino acid sequence is
\begin{equation}
\begin{matrix} {\tt  \ W \ M \ E \ W \ D \ R \ E \ I \ N \ N \ Y \ T \ S \ L \ I \ H \ S  } \\
{\tt  L \ I \ E \ E \ S \ Q \ N \ Q \ Q \ E \ K \ N \ E \ Q \ E \ L \ L} \end{matrix}
\la{lets}
\end{equation}
These amino acids are assigned the numbers 628-661 in the PDB entry of 1AIK.

\subsection{All-atom simulations}

We have used the molecular dynamics package GROMACS 4.6.3 \cite{Hess-2008}. We have analysed in detail a number of 
80 ns long trajectories, with the crystallographic PDB conformation as the initial condition. We have chosen
the length of the trajectories by inspecting, when major structural deformations take place.
We have employed  three different force fields,  to eliminate force-field based artifacts.  These are 
the united-atom force field GROMOS53a6, and the all-atom force fields CHARMM27 and OPLS/AA.  

The 1AIK subchain that we have investigated in detail, 
consists of 34 amino-acid residues, with PDB numbers 628-661.
There are 16200 atoms in the entire system that
we have simulated, including the solvent. The simulation box has dimensions $47 \times 47 \times 74$ 
\AA$^3$. This ensures that there is a  2 nm minimal distance between the protein atoms and the box walls, 
with periodic boundary conditions. 

We have described the  solvent using the SPC water model \cite{Berendsen-1981}. We have neutralised the system at a 
salt concentration of 0.15 mol/l. We have used steepest-descent for initial energy minimisation. The system was warmed up to 290 K by a simulated annealing in a 100 ps position-restraint simulation. We have chosen this 
relatively low temperature value for a better control of random thermal noise but without forgoing the underlying
physical phenomena.
For temperature control we have employed the Berendsen-thermostat with a time constant 0.1 ps, and for pressure coupling --- the Berendsen-barostat with a pressure set to 1 bar and a time constant 0.5 ps. Constraints on all bonds were imposed with the LINCS algorithm \cite{Hess-1997}.  We have used the particle mesh Ewald (PME) method \cite{Darden-1993} to 
compute the long-range electrostatic interactions, with van der Waals and Coulomb cutoff radii of 0.9 nm. 
For the 80 ns production run with a time step of 2 fs, that we analyse here in detail,   
we have changed the thermostat to v-rescale and the barostat 
to Parrinello-Rahman, keeping the initial time constants, to ensure
the generation of a proper canonical ensemble \cite{Hess-2008}. We have recorded the coordinates  every 20 ps, which gives rise 
to 4000 frames that form the basis for our analysis.

\subsection{Protein geometry}

We have introduced, employed and developed a number of topological tools and analytic techniques to analyse and interpret
the results of our GROMACS simulations.
  
\subsubsection{Discrete Frenet equation}

We monitor the evolution of the protein geometry using 
Frenet frames which are based on the backbone C$\alpha$ atoms \cite{Hu-2011-a}. The framing
depends {\it only} on the C$\alpha$ atom  coordinates ${\bf r}_i$, where $i=0,...,N$ labels  the residues and $N= 33$ in the case of 1AIK.
At a given ${\bf r}_i$ the frame consists of the unit backbone tangent ($\mathbf t_i$), 
binormal ($\mathbf b_i$) and normal ($\mathbf n_i$) vectors, defined as follows,
\begin{equation}
\mathbf t_i = \frac{ {\bf r}_{i+1} - {\bf r}_i  }{ |  {\bf r}_{i+1} - {\bf r}_i | }
\la{eq:t}
\end{equation}
\begin{equation}
\mathbf b_i = \frac{ {\mathbf t}_{i-1} \times {\mathbf t}_i  }{  |  {\mathbf t}_{i-1} \times {\mathbf t}_i  | }
\la{eq:b}
\end{equation}
 \begin{equation}
\mathbf n_i = \mathbf b_i \times \mathbf t_i ~~~~
\la{eq:n}
\end{equation}
Our aim is to identify and isolate the collective multi-atom motions that 
drive the protein folding process, from the background of the various 
random fluctuations. We expect that such coherent motions and oscillations 
have characteristic time scales, which are much longer than the period of an individual
atom covalent bond oscillation.
In average, over the relevant time scales,  
the distance between two consecutive C$\alpha$ atoms can then be taken to be  
nearly constant, and equal to
\begin{equation}
|{\mathbf r}_{i+1}-{\mathbf r}_i|\approx 3.8 \ {\rm \AA}
\la{eq:aveva}
\end{equation} 
Thus, at relevant time scales, the backbone dynamics
can be described entirely in terms of the virtual backbone bond and torsion angles $\kappa_i$ and $\tau_i$, as the {\it complete structural order parameters} \cite{Hinsen-2013,Niemi-2014}. These angles are 
defined as follows,
\begin{eqnarray}
\kappa_{i+1,i} \ \equiv \  \kappa_i \  & = &  \arccos (  \mathbf t_{i+1} \cdot \mathbf t_i  ) 
\la{eq:bond}\\
\tau_{i+1,i} \ \equiv \ \tau_{i} \  & =  & \omega \, \arccos (\mathbf  b_{i+1} \cdot \mathbf b_i )
\la{eq:torsion}
\end{eqnarray}
where
\begin{equation}
\omega = {\rm sign}[(\mathbf b_{i-1}\times \mathbf b_i)\cdot\mathbf t_i] 
\la{eq:s}
\end{equation}
Conversely,  the frame vectors (\ref{eq:t})-(\ref{eq:n}) can be
expressed in terms of these two order parameters % bond angles $\kappa$ and torsion angles $\tau$, 
iteratively, using the {\it discrete Frenet equation}  \cite{Hu-2011-a}
\begin{equation}
\left( \begin{matrix} {\bf n}_{i+1} \\  {\bf b }_{i+1} \\ {\bf t}_{i+1} \end{matrix} \right)
= 
\left( \begin{matrix} \cos\kappa \cos \tau & \cos\kappa \sin\tau & -\sin\kappa \\
-\sin\tau & \cos\kappa & 0 \\
\sin\kappa \cos\tau & \sin\kappa \sin\tau & \cos\kappa \end{matrix}\right)_{\hskip -0.1cm i+1 , i}
\left( \begin{matrix} {\bf n}_{i} \\  {\bf b }_{i} \\ {\bf t}_{i} \end{matrix} \right) 
\la{eq:DFE2}
\end{equation}
and the C$\alpha$ backbone is calculated from
\begin{equation}
\mathbf r_k = \sum_{i=0}^{k-1} |\mathbf r_{i+1} - \mathbf r_i | \cdot \mathbf t_i
\la{eq:dffe}
\end{equation}
Unlike the tangent vector
$\mathbf t_i$, the normal and binormal vectors $\mathbf n_i$ and $ \mathbf b_i$ do not appear 
in equation (\ref{eq:dffe}). Thus, if we rotate these two vectors simultaneously around 
the vector $\mathbf t_i$, the C$\alpha$ geometry remains intact and
only the way how it is framed changes.
In particular, rotation by $\pi$ constitutes the discrete $\mathbb Z_2$ 
%%% added 
gauge 
%%%%
transformation 
\begin{equation}
\begin{matrix}
\ \ \ \ \ \ \ \ \ \kappa_{i }  & \to &  \hskip -2.5cm \kappa_{i} - \pi  \\
\ \ \ \ \ \ \ \ \ \tau_{k} & \to  &  - \ \tau_{k} \ \ \ \hskip 1.0cm  {\rm for \ \ all} \ \  k \geq i 
\end{matrix}
\la{eq:dsgau}
\end{equation}
This transformation has been previously used extensively, to analyse protein loop structure
\cite{Molkenthin-2011,Krokhotin-2012-a,Krokhotin-2014-a,Sieradzan-2014,Hu-2011-a,Hinsen-2013,Niemi-2014,Chernodub-2010,Hu-2011-b,Krokhotin-2011}. 
It will also be used in the sequel.

%%%%%%%%%%%%%%%%%
%%%%%%%%%%%%%%%%%%
%%%%%%%%%%%%%%%%%
%%%%%%%%%%%%%%%%%

\subsubsection{Heisenberg spin variables}

According to  (\ref{eq:dffe}) the entire C$\alpha$ backbone geometry is determined by the tangent vectors ${\bf t}_i$. Thus, 
following \cite{Lundgren-2012-a,Lundgren-2012-b,Peng-2014} we may
visualise the backbone geometry in terms of these vectors: We
take the base of $\mathbf t_i$ to be at the location $\mathbf r_{i}$  of the $i^{th}$ 
C$\alpha$ atom.  We identify the tip of $\mathbf t_i$  as a point on 
the surface  of a unit two-sphere $\mathbb S^2_i$ that is centered at the point $\mathbf r_{i}$. 
We orient the coordinate system on the sphere so that the north-pole coincides with the
tip of $\mathbf t_i$. Thus,  the north-pole is always in the direction of the next C$\alpha$, 
which is at the site $\mathbf r_{i+1}$. 
%The bond  angle $\kappa$ measures 
%the latitude of $\mathbb S_i^2$ from the north pole, and the torsion angle $\tau$  measures  the
%longitude starting from the great circle that passes both through the north pole 
%and through the tip of the binormal vector $\mathbf b_i$. 

We proceed to characterise the direction  
of the next tangent vector $\mathbf t_{i+1}$ {\it i.e.} the 
direction from $\mathbf r_{i+1}$ towards the C$\alpha$ atom at site $\mathbf r_{i+2}$,  in
terms of the longitude and latitude angles of the $i^{th}$ two-sphere $\mathbb S_i^2$. For this, 
we translate the center of $\mathbb S^2_i$  from $\mathbf r_i$  
towards its north-pole, and all the way to the location $\mathbf r_{i+1}$ of the $(i+1)^{th}$ C$\alpha$ atom, 
without introducing any rotation of the sphere. We  then record the direction of $\mathbf t_{i+1}$  
as a point on the surface of the translated $\mathbb S^2_i$.  This defines the 
coordinate values ($\kappa_i, \tau_i$), that determine how the backbone chain  turns 
at site $\mathbf r_{i+1}$, to reach  the  $(i+2)^{th}$ central C$\alpha$ atom at the point $\mathbf r_{i+2}$:
The angle $\kappa_i$ measures the latitude of $\mathbf t_{i+1}$ on the translated two-sphere 
$\mathbb S^2_i$,  from its north pole. The angle $\tau_i$ measures the longitude 
of $\mathbf t_{i+1}$, starting from the great circle that passes both through the north pole and 
through the tip of the binormal vector $\mathbf b_i$. 
 
When we repeat the above procedure for all C$\alpha$ atoms, we obtain a ($\kappa, \tau$) 
distribution that characterises the overall geometry of a protein backbone.  
For a visualisation of this distribution  we employ  the geometry of a stereographically 
projected  two-sphere: We  project the $(\kappa, \tau)$ coordinates from the south pole
to the tangent plane of the north pole, of the two-sphere. If ($x,y$) are the coordinates of this 
tangent plane the projection is defined by 
\begin{equation}
x + iy =  \tan  (\frac{\kappa}{2}) \cdot e^{- i \tau}
\la{stere}
\end{equation}

When we perform the projection for all C$\alpha$ atoms in all crystallographic protein structures in PDB that
have been measured with resolution better than 2.0 \AA,  
we arrive at the statistical angular  distribution that we show  in Figure \ref{fig-2}. 
%
%
%
%          figure-2b
%
%
%
\begin{figure}
\centering
{\begin{minipage}[t]{.45\textwidth}
  \raggedright
  \includegraphics[trim = 0mm 65mm 0mm 70mm, width=0.9\textwidth,  angle=0]{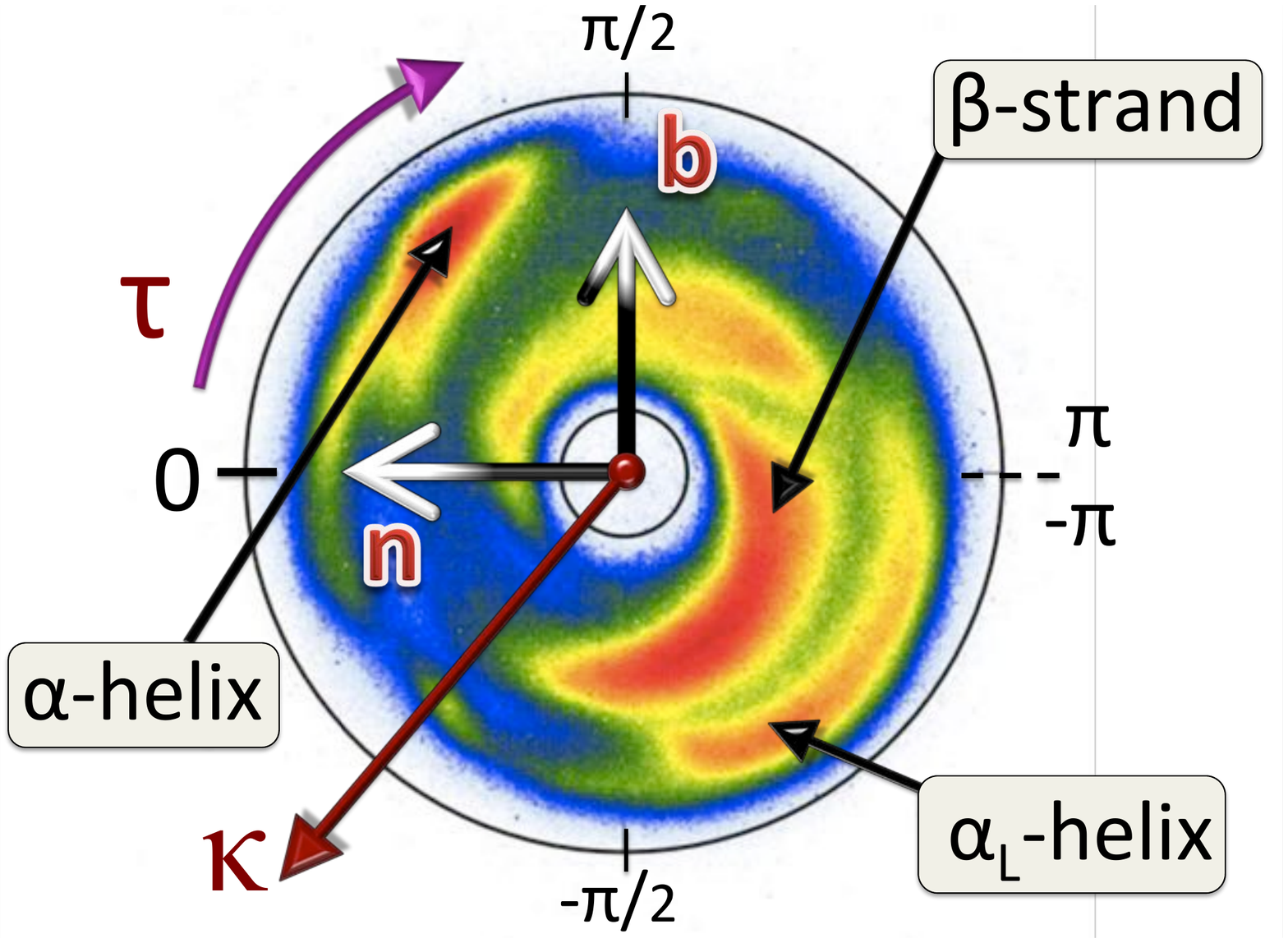}
  \caption{The distribution of ($\kappa,\tau$) values in all PDB structures with better than 2.0 \AA~
  resolution, on the stereographically projected two-sphere, with a rainbow encoding of 
  the number of entries (red corresponding to the largest number). The locations of the major regular secondary structures are identified.}
  \la{fig-2}
\end{minipage}}
\end{figure}
It  is the  landscape for the shape of the protein backbones from the crystallographic data in
PDB. By the way it is obtained, the crystallographic protein structure should be very close to a
stationary minimum of the ensuing Gibbs free energy. Thus the figure \ref{fig-2} should be 
the collective landscape of stationary, minimum-energy protein structures.

We observe that the PDB data is concentrated in an annulus which is roughly between the 
circles $ \kappa_{in} \approx 1$ and $\kappa_{out} \approx  \pi/2$. 
The exterior of the 
annulus $\kappa > \kappa_{out}$ is an excluded region, the ensuing conformations are subject to steric clashes. The  
interior $\kappa <  \kappa_{in}$ is  sterically allowed but in practice excluded in PDB structures.
Note that regular structures such as $\alpha$-helices and  $\beta$-strands are distinguished
as highly localised regions in the figure  \ref{fig-2}, with 
\[
(\kappa,\tau)_\alpha \approx (1.57,0.87) \sim (\frac{\pi}{2}, 1)
\]
for $\alpha$-helices and 
\[
(\kappa,\tau)_\beta \approx (1,-2.9) \sim (1, \pm\pi)
\]
for $\beta$-strands. %Loops are trajectories with variable ($\kappa,\tau$), they connect different regions in Figure  \ref{fig-2}.  
Different regions in Figure  \ref{fig-2} can be connected by loops, which can be considered as trajectories along the variables ($\kappa,\tau$).
We have found that 
loops have the tendency to encircle the inner circle. 
In figure \ref{fig-3} a) we show, as an example, a generic
loop that connects the right-handed $\alpha$-helical region with the $\beta$-stranded region. 
%
%
%          figure-3b
%
%
%
\begin{figure}
\centering
{\begin{minipage}[h]{.45\textwidth}
  \raggedright
  \includegraphics[trim = 0mm 5mm 0mm -10mm, width=0.9\textwidth,  angle=0]{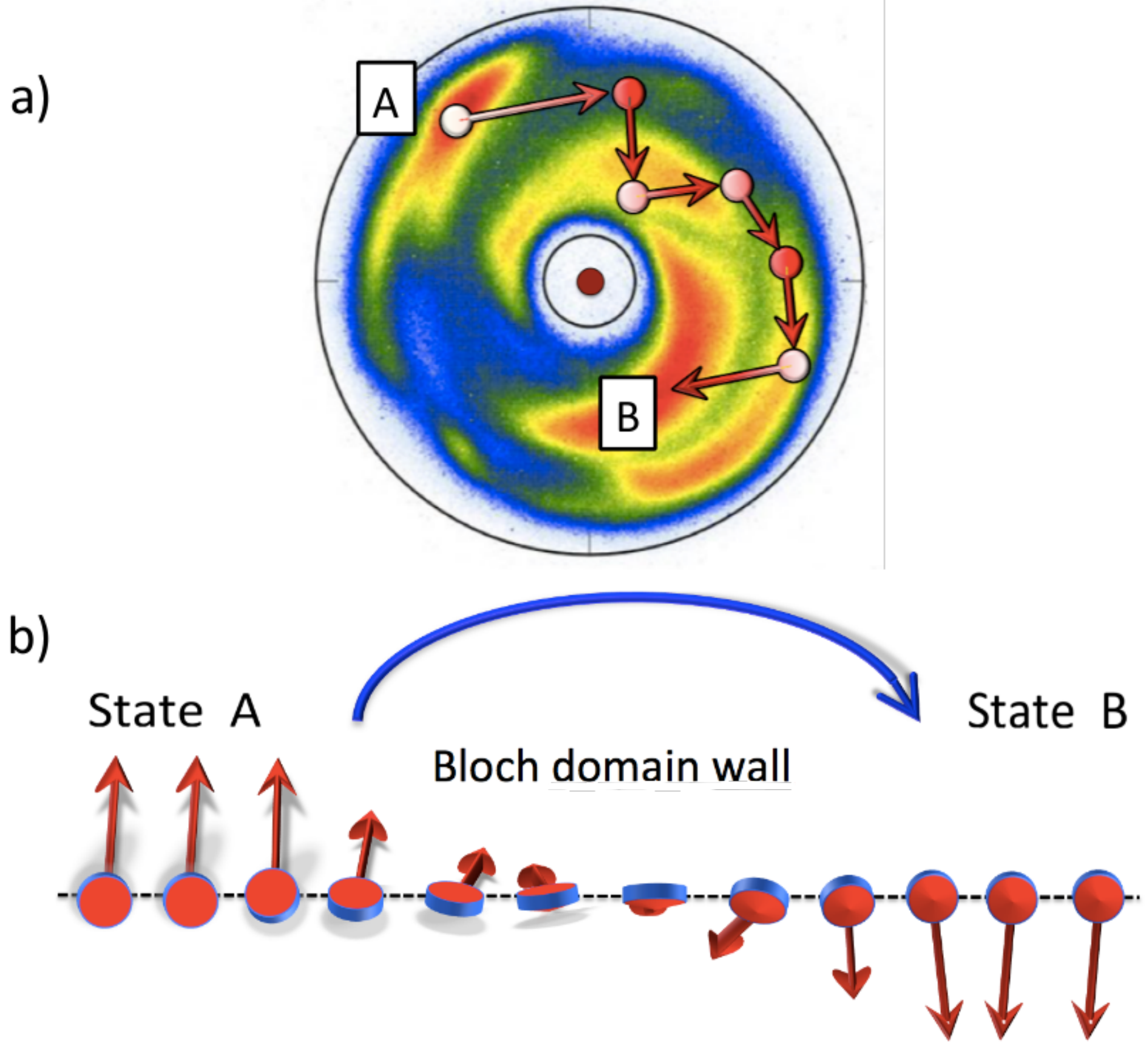}
  \caption{a) A generic loop is a trajectory on the stereographically projected ($\kappa,\tau$) sphere, that connects 
  a region corresponding to a regular secondary structure (here A) to another one (here B). b) In terms of the variable (\ref{si}),
  a loop becomes  a Bloch domain wall that interpolates between ground states A and B, along a 
  Heisenberg spin chain.}
  \la{fig-3}
\end{minipage}}
\end{figure}

To describe a  backbone segment analytically,  we combine its C$\alpha$'s bond and torsion angles 
into the three component unit vectors 
\begin{equation}
\mathbf s_i \ = \ \left( \begin{matrix} \cos \tau_i \sin\kappa_i  \\ \sin\tau_i \sin\kappa_i \\ \cos\kappa_i \end{matrix}
\right)
\la{si}
\end{equation}
We interpret these vectors as the local order parameters along an imaginary {\it linear} one dimensional 
Heisenberg spin chain, labeled by the index $i$.  This converts the  C$\alpha$  %backbone 
geometry into 
a configuration along a linear Heisenberg chain in a one-to-one manner:
%; the spin variable takes values in the 
%annulus  $\kappa_{in} < \kappa <  \kappa_{out}$ of the two-sphere $\mathbb S^2$. 
In Figure \ref{fig-3}(b) we have sketched how the (generic) trajectory shown in Figure  \ref{fig-3}(a) appears, 
figuratively, in terms of such a Heisenberg spin chain configuration.
 
Since the spin variable (\ref{si}) takes values only in the annulus  $\kappa_{in} < \kappa <  \kappa_{out}$ of 
the two-sphere $\mathbb S^2$, it is apparent that a loop can be {\it de facto} identified as a  domain wall akin the Bloch 
wall along a Heisenberg chain. The loop then
interpolates between the two different regular secondary
structures, denoted by state A and state B respectively, as shown in the figures 3. 

\subsubsection{Residues and spin chains} 

The amino acid side chains can be similarly interpreted in terms of a one dimensional 
linear spin chain.  In fact, there are several
ways to identify the spin chain variable. Here we utilise the
directional vector that points from the C$\alpha$ atom at $\mathbf r_i$ towards 
the ensuing C$\beta$ atom,  located at $\mathbf r^\beta_i$.  This vector can be introduced 
for all amino acids except glycine (G); note that there is no glycine in (\ref{lets}). 

We start with the unit vector
\begin{equation}
\mathbf u^\beta_i \ = \  \frac{ \mathbf r_i^\beta - \mathbf r_i}{  |\mathbf r_i^\beta - \mathbf r_i |}
\la{sideO3}
\end{equation}
We recall  the C$\alpha$ based discrete Frenet framing with the coordinates
($\kappa_i, \tau_i$) and represent (\ref{sideO3}) as 
three component unit vectors in this coordinate system,
\begin{equation}
\mathbf u^\beta_i \ \to \ \hat\sigma_i \ = \ \left( \begin{matrix} \cos \tau_i^\beta \sin \kappa_i^\beta \\
\sin \tau_i^\beta \sin \kappa_i^\beta \\ \cos \kappa_i^\beta \end{matrix} \right)
\la{cbetai}
\end{equation}
Here ($\kappa_i^\beta, \tau_i^\beta$) are the spherical coordinates of the $i^{th}$ C$\beta$ atom,
on the surface of the C$\alpha$ centered two-sphere $\mathbb S^2_i$.
In Figure \ref{fig-4} we show the distribution of the vectors (\ref{cbetai}) on the surface of the two-sphere,
for all those crystallographic PDB structures that have been measured with resolution better than 1.0 \AA.
Note that the sphere is the same as in figures \ref{fig-2} and \ref{fig-3}(a) but  now there is no 
stereographic projection.
%
%
%          figure-4b
%
%
%
%\end{document}
\begin{figure}
\centering
{\begin{minipage}[h]{.45\textwidth}
  %\raggedright
  \includegraphics[trim = 0mm 5mm 20mm -10mm, width=0.9\textwidth,  angle=0]{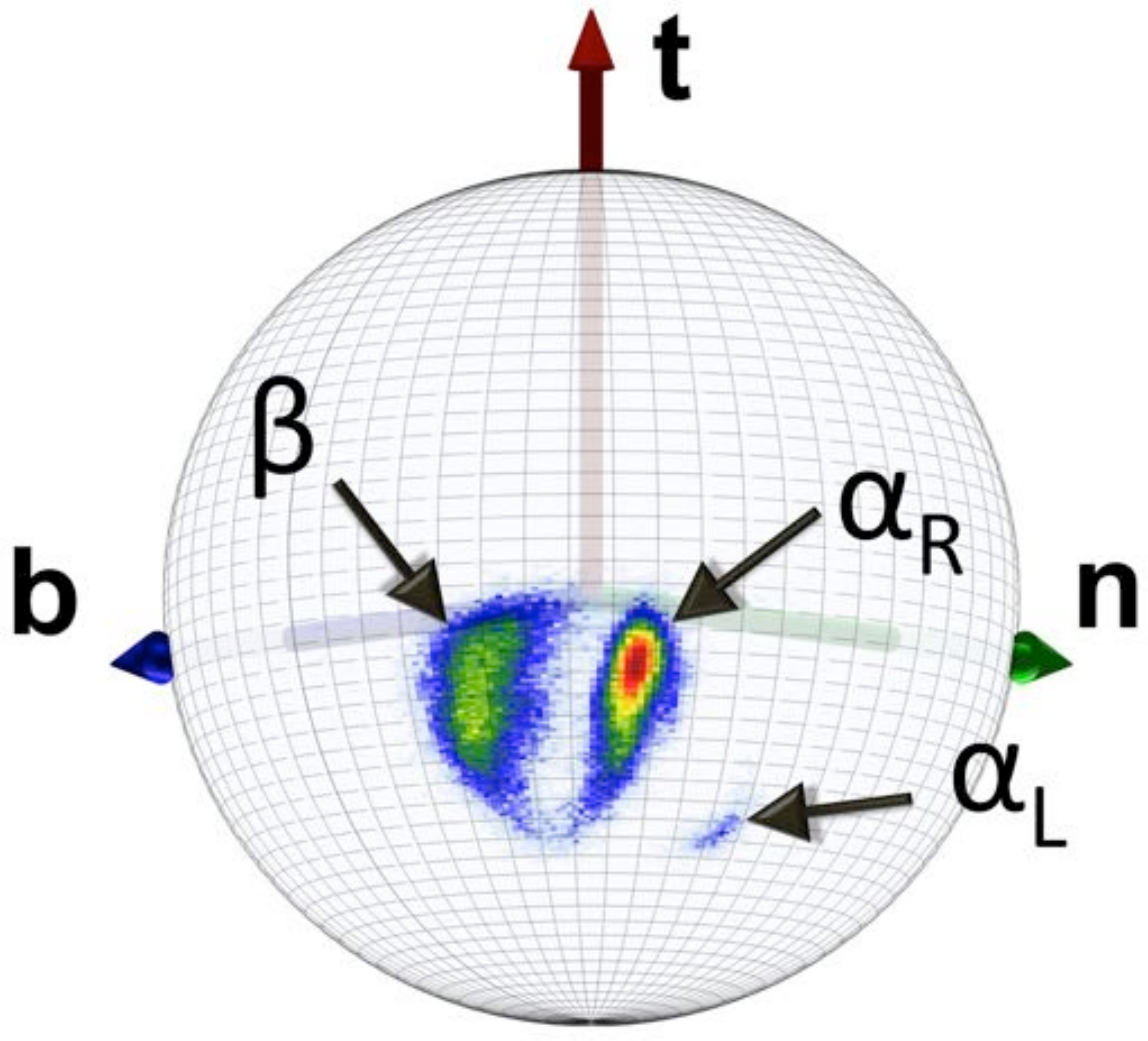}
  \caption{ C$\beta$ distribution in the corresponding C$\alpha$ centered discrete Frenet frames for
  all structures in PDB with resolution better than 1.0 \AA. The regions corresponding to $\alpha$-helices ($\alpha_R$),
  $\beta$-strands ($\beta$), left-handed $\alpha$-helices ($\alpha_L$)  are identified, the rest are (mostly) loops. }
  \la{fig-4}
\end{minipage}}
\end{figure}

We can interpret the distribution in Figure \ref{fig-4} as the C$\beta$ landscape of stationary
folded protein structures with minimum Gibbs energy. The highly localized character of the distribution 
shows that there is a very strong correlation between the C$\alpha$ (backbone)
geometry and the C$\beta$ (side chain) geometry. 
Accordingly,  the ground state structures of the corresponding Heisenberg spin chain Hamiltonians
must be very similar.

We proceed to introduce a set of  O(2) spin variables for the side-chain C$\beta$.  For this we
define the projection of (\ref{sideO3}) onto the normal plane at the position of the $i^{th}$ C$\alpha$, 
\[
\mathbf u_i \ = \ \frac{ \mathbf u_i^\beta - (\mathbf u_i^\beta \cdot \mathbf t_i) \mathbf t_i } 
{ | \mathbf u_i^\beta - (\mathbf u_i^\beta \cdot \mathbf t_i) \mathbf t_i  |}
\]
For the next C$\beta$ along the chain, we introduce similarly the vector $\mathbf u_{i+1}$ and compute
its projection onto the {\it same} normal plane --- at the position of the $i^{th}$ C$\alpha$, 
\[
\mathbf v_{i} \ = \ \frac{ \mathbf u_{i+1} ^\beta - (\mathbf u_{i+1}^\beta \cdot \mathbf t_i) \mathbf t_i } 
{ | \mathbf u_{i+1}^\beta - (\mathbf u_{i+1}^\beta \cdot \mathbf t_i) \mathbf t_i  |}
\]
We then define the relative angle $\eta_i$,
\begin{equation}
\cos\eta_i \ = \ \mathbf u_i  \cdot \mathbf v_i 
\la{etai} 
\end{equation}
As shown in Figure \ref{fig-5},  $\eta_i$ is the dihedral angle
\begin{equation}
\eta_i \ := \ C\beta(i) \ - \ C\alpha(i) \ - \ C\alpha(i+1) \ - \ C\beta(i+1)
\la{etai2}
\end{equation}
We note that the construction resembles that of Newman projection
in stereochemistry.
%
%
%          figure-5b
%
%
%
\begin{figure}
\centering
{\begin{minipage}[h]{.45\textwidth}
  %\raggedright
  \includegraphics[trim = 0mm 5mm 20mm -10mm, width=0.7\textwidth,  angle=0]{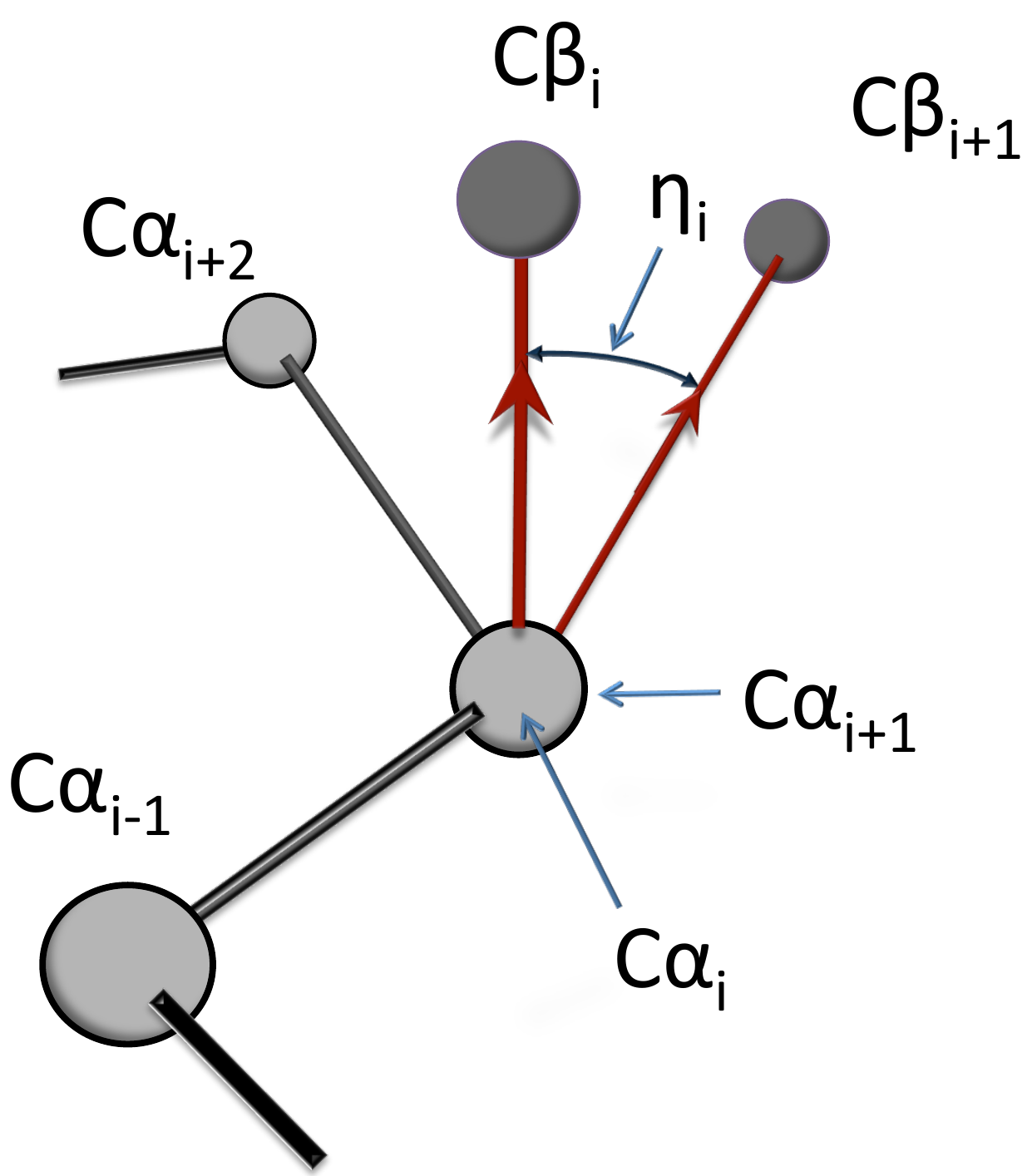}
  \caption{The angle $\eta_i$ in (\ref{etai}) is defined as the angle between 
  the projections of the vectors $\mathbf u_i$ and $\mathbf u_{i+1}$, connecting the $i^{th}$ C$\alpha$
  and C$\beta$, and the $(i+1)^{st}$ C$\alpha$ and C$\beta$ %, when both vectors are projected 
  on the normal plane of $\mathbf t_i$. Note that in the figure the $i^{th}$ C$\alpha$
  is in front of (on top of) the $(i+1)^{st}$ C$\alpha$. }
  \la{fig-5}
\end{minipage}}
\end{figure}

In analogy with Figure \ref{fig-3}(b) we identify the variable $\eta_i$ as an order 
parameter for a linear O(2) XY spin chain, 
\begin{equation}
\mathbf  m_i \ = \ \left( \begin{matrix} \cos \eta_i \\ \sin \eta_i \end{matrix} \right)
\la{mi}
\end{equation}
 
Like the Heisenberg model, the XY model supports domain 
walls  that interpolate between two configurations where the order parameter 
has different constant values. The domain
wall of the XY model is akin the Bloch domain wall of a Heisenberg model. 
Figure \ref{fig-4} shows that  the ground state 
structure of the side chain XY model is closely related to that of the backbone 
Heisenberg model, in the case of crystallographic PDB protein structures.

\subsubsection{Folding indices}

The formation, evolution and structure of a loop along a folding protein  can be monitored
in terms of topologically determined folding indices. Here we are interested in two particular
examples of folding indices,
one that relates to the backbone  geometry and another one that relates to the side chain geometry.

In the case of a Heisenberg spin chain,  there is 
a topological index akin a winding number that characterises and classifies its 
Bloch domain walls. For the C$\alpha$ Bloch wall  
shown in figure \ref{fig-3} a), b) this topological index 
counts the net number of  times the corresponding trajectory encircles
around the annulus in the figure {\it i.e.} around the 
north-pole of the two-sphere. We remind that due to steric constraints, the Heisenberg variable (\ref{si}) 
takes values in  the annulus shown in figures \ref{fig-2}, \ref{fig-3} a). We also recall  
that for the first homotopy class  of a circle $\pi_1 (\mathbb S^1) \simeq \mathbb Z$ which justifies the
introduction of a topological concept.

Analytically, we may assign to each loop, and more generally to a backbone segment, between residues $n_1$ and $n_2$
the following  folding index  $Ind_f$ \cite{Lundgren-2013},  
\begin{equation}
Ind_f = \left[\, \Gamma\, \right] 
\la{eq:Gamma}
\end{equation}
where
\begin{equation}
\Gamma = \frac{1}{\pi}\sum\limits_{i=n_1+2}^{n_2-2} \begin{cases}
     \tau_{i}-\tau_{i-1}-2\pi & {\rm if}\   \tau_{i}-\tau_{i-1} > \pi\\
     \tau_{i}-\tau_{i-1}+2\pi & {\rm if}\  \tau_{i}-\tau_{i-1} < -\pi\\
     \tau_{i}-\tau_{i-1} & {\rm otherwise}
\end{cases}
\la{eq:foldind}
\end{equation}
Here $[x]$ denotes the integer part of $x$. Note that $\Gamma$ is the total rotation angle (in radians)
that the projections of the C$\alpha$ atoms of the consecutive loop (segment) residues make around the north pole. The
$n_1, \ n_2$ label the first and last residue of the loop. Commonly these are the last {\it resp.}  first residues in
the preceding and in the following regular secondary structures.
The folding index is a positive integer when the rotation is counterclockwise, and a negative integer 
when the rotation is clockwise.  The folding index can be used to classify individual loop structures and backbone
segments, even entire protein backbones \cite{Lundgren-2013}.  
Note that the folding index is normalised so that it is equal to twice the number of times 
the vector in Figure  \ref{fig-3}(b) rotates around its axis, when the spin structure traverses a domain wall {\it i.e.} it assigns 
an even integer to the  $\pi_1 (\mathbb S^1) \simeq \mathbb Z$ winding number, in the case of a closed trajectory.

For example, for the trajectory shown in figures \ref{fig-3} the folding index has the value -1. 
For a loop connecting an $\alpha$-helix and a $\beta$-strand, the folding index is generically an odd integer. For
a loop connecting two $\alpha$-helices, or two $\beta$-strands, the folding index is generically an even integer.

In the case of the side chains, we utilise the C$\beta$
angular XY spin variable  (\ref{etai}) to define a similar topological folding index \cite{Sieradzan-2014}. For this, 
we first choose a reference residue, {\it e.g.} the $n^{th}$ residue along the backbone.
Starting with this reference residue, we then evaluate 
the accumulated {\it total} angle $\hat \eta_m$ over a segment with $m-n$ residues, 
\begin{equation}
\hat \eta_m= \sum_{k=n}^m \eta_k
\la{eq:etaprime}
\end{equation}
and we define the ensuing index by
\begin{equation}
Ind_m \ = \ [ \frac{\hat\eta_m}{\pi}]
\la{eq:etaprime2}
\end{equation}
Again, the index acquires its topological justification from the fact that 
$\pi_1(\mathbb S^1) \simeq \mathbb Z$. The dihedral $\eta_k$, the 
accumulated total angle (\ref{eq:etaprime}) and
the ensuing index (\ref{eq:etaprime2}) can all be used to study and classify loop structures, protein segments,  
and entire proteins.

\subsubsection{Landau free energy} 
\la{sect:landau}

A generic all-atom molecular dynamics simulation of a folding protein contains a wide range of intermediate 
conformations. Some of these are essential for the correct folding pathway, while some are merely 
random transients with no inherent relevance to the folding process {\it per se}.
In order to identify the relevant conformational processes, we need systematic 
methods that smooth over and weave out the irrelevant random fluctuations. 
For this we recall the standard Wilsonian universality arguments \cite{Kadanoff-1966,Wilson-1971}, 
to deduce the form of the Landau free energy that emerges from the thermodynamical Gibbs free energy. 
In the present context, the derivation is based on the following two assumptions \cite{Niemi-2014}:

$\bullet$ We assume that the characteristic length scales that are associated with  spatial 
variations and deformations along  the protein backbone around its thermal equilibrium 
configuration, are large in comparison to the covalent bond lengths. This presumes 
that there  are no abrupt edges but only gradual slowly varying bends and twists 
along the backbone. From Figures \ref{fig-2} and \ref{fig-4} we conclude that  
the steric constraints between the backbone and the side chain atoms act as 
powerful inhibitors of sharp, edgy motions. 

$\bullet$  We also assume that the individual C$\alpha$ virtual bond length oscillations 
have characteristic time scales, which are very short in comparison to the time scale which characterise
a folding process. The characteristic time scale of a random covalent bond oscillation is around ten  
femtoseconds while in our simulations we  record the individual atomic coordinates every 20 picosecond.  
We have tested that conformational changes which take place over shorter time scales, do not affect our
conclusions. Accordingly, we may adopt  (\ref{eq:aveva}) as the (time averaged) value for all the 
nearest neighbour C$\alpha$-C$\alpha$ distances.

It has been shown that  in the case of crystallographic PDB structures,
the bond and torsion angles ($\kappa_i, \tau_i$) form a {\it complete set 
of structural order parameters} \cite{Hinsen-2013}. Accordingly,
in the vicinity of a Gibbs free energy minimum we may expand the 
free energy in terms of these angles. For this we consider the response of the  interatomic 
distances to variations in these angles, with
\[
r_{ab} \ = \  r_{ab} (\kappa_i, \tau_i)
\]
where $r_{ab}$ is the distance between {\it any} two C$\alpha$ atoms $a$ and $b$
along the backbone.
 
Suppose that at a given local extremum of the  free energy, the C$\alpha$ bond and torsion angles have the
equilibrium  values
\[
(\kappa_i , \tau_i) \ = \ (\kappa_{i0},  \tau_{i0})
\]
We then consider a non-equilibrium conformation where the ($\kappa_i, \tau_i$) deviate from  these
extremum values. We denote the deviations by
\begin{equation}
\begin{matrix} 
\Delta \kappa_i & = & \kappa_i - \kappa_{i0} \\
\Delta \tau_i & = & \tau_i - \tau_{i0}
\end{matrix}
\la{devi}
\end{equation}
When the deviations are slowly varying in space, {\it i.e.} (\ref{devi}) are small, we may Taylor expand the Gibbs free energy  
around the extremum, 
\[
G( r_{\alpha\beta} ) \ \equiv \
G\left[  r_{\alpha\beta} (\kappa_i, \tau_i)\right] \ =
\]
\[ 
 G(\kappa_{i0}, \tau_{i0}) + \sum\limits_k \left \{ \, 
\frac{\partial G}{\partial \kappa_k}_{|0}\! \Delta \kappa_k 
+ \frac{\partial G}{\partial \tau_k}_{|0}\! \Delta\tau_k \, \right \} +
\]
\[
+  \sum\limits_{k,l} \left\{ \, \frac{1}{2}
\frac{\partial^2 G}{\partial \kappa_k \partial \kappa_l }_{|0}\! \Delta \kappa_k  \Delta \kappa_l
 +
\frac{1}{2}
\frac{\partial^2 G}{\partial \tau_k \partial \tau_l }_{|0}\! \Delta \tau_k  \Delta \tau_l \right.
\]
\[
\left. + \frac{\partial^2 G}{\partial \kappa_k \tau_l}_{|0}\! \Delta\kappa_k \Delta \tau_l  
\, \right \} + \mathcal O (\Delta^3)
\]
The first term in the expansion evaluates the free energy at the extremum. Since ($\kappa_{i0}, \tau_{i0}$)
correspond to the extremum, the second term vanishes. Thus we are left with 
the following expansion of the averaged 
free energy,
\[
G(\kappa_i, \tau_i) \ = \ G(\kappa_{i0}, \tau_{i0}) \
\]
\[
+  \ \sum\limits_{k,l} \left \{ \, \frac{1}{2}
\frac{\partial^2 G}{\partial \kappa_k \partial \kappa_l }_{|0}\! \Delta \kappa_k  \Delta \kappa_l
+
\frac{1}{2}
\frac{\partial^2 G}{\partial \tau_k \partial \tau_l }_{|0}\! \Delta \tau_k  \Delta \tau_l
\, \right.
\]
\begin{equation}
\left. + \frac{\partial^2 G}{\partial \kappa_k \tau_l}_{|0}\! \Delta\kappa_k \Delta \tau_l
\right \} + \dots 
\la{Fene}
\end{equation}
When the characteristic length scale of spatial deformations around a minimum energy configuration is large
in comparison to a covalent bond length, we may  {\it re-arrange} 
the expansion (\ref{Fene}) in terms of the differences in the angles as follows:
first come local terms, then come terms that connect the nearest neighbours, 
then come terms that connect the next-to-nearest neighbours,  and so forth.
When we re-order the expansion (\ref{Fene}) in this manner  and demand that 
the free energy is invariant under local rotations in the ($\mathbf b_i, \mathbf n_i$)-plane, we conclude
\cite{Niemi-2003,Danielsson-2010,Niemi-2014} 
that to the leading order the expansion of the Gibbs free energy  {\it necessarily} coincides  
with the energy of the following discrete nonlinear Schr\"odinger equation \cite{Faddeev-1987,Ablowitz-2004,Molkenthin-2011}
\[
F  =  \sum\limits_{i=1}^N
\left \{  \lambda\, (\kappa_{i}^2 - m^2)^2  + \frac{q}{2} \, \kappa_{i}^2 \tau_{i}^2   
- p \,  \tau_{i}   +  \frac{r}{2}  \tau^2_{i} + \dots
\right\} 
\]
\begin{equation}
+ \sum\limits_{i=1}^{N-1}  ( \kappa_{i+1} - \kappa_{i})^2   \ + \dots
\la{E1old}
\end{equation}
\begin{equation}
\equiv \ V_{pot} [\kappa,\tau] + \sum\limits_{i=1}^{N-1}  ( \kappa_{i+1} - \kappa_{i})^2 
\la{E2old}
\end{equation}  
This functional form of the free energy is simply the most general Landau free energy 
that one can write down using the available variables ($\kappa_i,\tau_i$), in a manner which
is consistent with the symmetry principle that a local
rotation of the ($\mathbf n_i, \mathbf b_i$) frames has no effect on the backbone geometry.
The corrections to (\ref{E1old}) include next-to-nearest neighbours 
couplings and so forth, which are higher order terms
from the point of view of our systematic expansion.  

We note that the expansion (\ref{E1old}) has the property that in 
continuum limit it yields the Coleman-Weinberg  derivative (low momentum)
expansion \cite{Coleman-1973}
\begin{equation}
F \ \rightarrow \ \int\limits_0^L ds \ \{ \ V(\phi) +A +  |(\partial_s + iA) \phi|^2 +  \dots \}
\la{coleman}
\end{equation}
where, following \cite{Niemi-2003,Danielsson-2010}, we have identified 
the bond angle with the complex scalar field $\kappa_i \to \phi(s)$ and the torsion angle with the U(1) gauge
field $\tau_i \to A(s)$, in the continuum limit. 

We {\it emphasize}  that the approximation (\ref{E1old}) is valid in the limit of slow spatial variations (low momentum).
That is, as long as there are 
no abrupt, sharp  edges but only gradual  bends and twists along 
the backbone. In particular, long range interactions 
are accounted for as long as they do not cause any localised 
sharp buckles along the backbone, and the angular variations respect the steric constraints.

The Wilsonian universality arguments are sufficient to conclude that in the limit of slowly varying backbone geometry
{\it any} complete all-atom force field can be approximated by the energy function (\ref{E1old}). The
parameters $\lambda$, $q$, $p$, $r$, and $m$ 
depend on the atomic level physical properties and the chemical 
microstructure of the protein and its environment. In principle, these parameters can 
be computed from this knowledge. But as always in the case of a Landau free energy,
it remains a challenge to compute these
parameters from the all-atom level.
 
\subsubsection{Spontaneous symmetry breaking and solitons} 
\la{sect:solitons}

The free energy (\ref{E1old}) relates to the DNLS energy function 
 \cite{Faddeev-1987,Ablowitz-2004,Molkenthin-2011}. The non-linear, quartic bond angle contribution 
 is the familiar double-well potential that gives rise to a spontaneous breakdown of
the $\mathbb Z_2$ symmetry
\[
\kappa_i \ \longleftrightarrow \ -\kappa_i
\]
The spontaneous breakdown of this discrete symmetry
is pivotal for the emergence  of a loop structure, in the case of proteins. 
It gives rise to a Bloch wave 
that interpolates between the two ground states $\kappa_i = \pm m$. 

More generally, the quartic potential admits a non-symmetric profile of the form 
\begin{equation}
U \approx \sum\limits_{{\rm C}\alpha} \frac{1}{2} k_0 (\kappa-a)^2(\kappa-b)^2 
\la{eq:corV}
\end{equation}
Here $a$ and $b$ are the positions of the minima of the quartic potential,  
and $k_0$ is a force constant. By carefully taking the continuum limit of the C$\alpha$ lattice, {\it i.e.} the limit where 
(\ref{eq:aveva}) becomes small, and by introducing a mass-scaled 
variable $\xi$  with $m$ representing the effective mass
of a residue, the pertinent DNLS equation becomes 
\begin{equation}
m \frac{d^2\xi}{ds^2} \ = \ - k_0 \, \xi \, (\xi^2 - c^2)
\la{eq:schr}
\end{equation}

\noindent where $s$ is the arc length parameter along the backbone. 
With $c=(a+b)/2$ and
\begin{equation}
m\xi  = \kappa - \frac{1}{2} (a+b)
\la{eq:xi}
\end{equation}
the solution of equation (\ref{eq:schr}) is 
\begin{equation}
\xi(s) = c \, \tanh \left[ c\sqrt{\frac{k_0}{2m}} (s-s_0) \right]
\la{kappaprof}
\end{equation}
where $s_0$ is the position of the inflection point, {\it a.k.a.} the center of a kink.
In terms of the original variables and parameters
\begin{equation}
\kappa (s) =    \frac{ \, b \, e^{ c \sqrt{ \frac{k_0}{2m} }  
(s-s_0) }   + a\,   e^{-  c \sqrt{ \frac{k_0}{2m} } \,  (s-s_0)}  }
{\cosh\left[c \sqrt { \frac{k_0}{2m} }(s-s_0)\right] }
\la{eq:cosh}
\end{equation}
This is known as the dark soliton solution of the nonlinear Schr\"odinger equation. It interpolates between
the asymptotic values which correspond to the (local) minima of the potential,
\begin{equation}
\kappa(s) \, \to \, \left\{ \begin{matrix}  \, a \ \ \ \ \ \ \ s\to-\infty  \\
\, b \ \  \ \ \ \ \ s\to+\infty 
\end{matrix} \right.
\la{asymp}
\end{equation}
In the case of a protein, the soliton describes the bond angle profile of a super-secondary structure such as 
($\alpha$-helix)-(loop)-($\beta$-strand) shown in 
figure \ref{fig-3}; the parameters have the values  $a \approx 1.5$ and $b \approx 1.1$ (radians) for the states $A$ and
$B$, shown in the figure.
%For example, the asymptotic parameter values $(\theta\approx\pi/2,\gamma\approx 1)$
%describe solitons that interpolate between two right-handed $\alpha$-helices, and 
%parameter values $(\theta\approx 1,\gamma\approx\pi)$ 
%describe solitons that interpolate between two $\beta$-strands.

In the case of a protein chain, the arc length $s$ becomes replaced by a
%of equations (\ref{eq:schr}--\ref{eq:cosh})
discrete variable which is equal to the position of the ensuing C$\alpha$  in the sequence. 
The variables $\kappa_i$ and $\tau_i$ are also mutually interacting, according to (\ref{E1old}).
The soliton is constructed as the minimum of $F$ in equation (\ref{E1old}) 
\cite{Molkenthin-2011,Chernodub-2010, Hu-2011-b,Niemi-2014}.  
It is the solution of a system of $2N-5$ nonlinear equations in
$2N-5$ unknowns, where $N$ is the number of residues. In order to obtain the solution, 
we first solve for $\tau_i$ in terms of $\kappa_i$,
\begin{equation}
\tau_{i} [\kappa] = \frac{p}{r + q\,\kappa^2_{i} } \ \equiv \ \frac{u }{1 + v \, \kappa^2_{i} } 
\la{tauk} 
\end{equation}

\noindent with $u=p/r$ and $v=q/r$.
By inserting equation (\ref{tauk}) into
equation (\ref{E1old}), the torsion angles $\tau$ are eliminated and we obtain a system of equations 
for the bond angles $\kappa$, 
\begin{equation}
\kappa_{i+1} = 2\kappa_i - \kappa_{i-1} + \frac{ d V_{pot}[\kappa]}{d\kappa_i^2} \kappa_i  \ \ \ \ \ (i=1,...,N)
\la{nlse}
\end{equation}
where $\kappa_0 = \kappa_{N+1}=0$
and 
\begin{equation}
V_{pot}[\kappa]  =  \frac{p}{r + q \, \kappa^2 } +  2 (1- \lambda m^2 ) \kappa^2
+ \lambda \, \kappa^4
\la{U}
\end{equation}
Here we recognize the discretised structure of equation  (\ref{eq:schr}).  
The difference is in the first term on the right-hand side in equation (\ref{U}). However, it turns out that in the
case of proteins, its effect is not that pronounced
as the effect of the other terms; it turns out that the first term is small 
in value when compared to the other two.

We can construct the profile of the dark soliton solution to equation (\ref{nlse}) 
numerically, by following the iterative procedure introduced in reference
\cite{Molkenthin-2011}; the explicit form of the solution is  until now unknown, 
in terms of elementary functions.
However,  we obtain an {\it excellent} approximation \cite{Hu-2011-b} by {\it naively} discretising the 
continuum dark nonlinear Schr\"odinger equation  soliton  (\ref{eq:cosh})
\begin{equation}
\kappa_i  =  \frac{ \mu_1
 \, \exp\left[ \sigma_{1} ( i-s)  \right] + \mu_{2} \, \exp\left[ - \sigma_{2} ( i-s)\right]}
{\exp\left[ \sigma_{1} ( i-s) \right] +  \exp\left[- \sigma_{2} ( i-s)\right]} 
\la{eq:bond2}
\end{equation}
%where $s$ is a parameter that determines the center of the soliton.
Here $\mu_{1,2} \in [0,\pi] \ {\rm mod}(2\pi)$ are parameters, which determine the amplitude of the variation
of $\kappa$ and the asymmetry of the inflection regions. The parameters $\sigma_1$ and $\sigma_2$ are
related to the inverse of the range of the inflection region. 
We remark that in the case of proteins, the values of $\mu_{1,2}$ are determined entirely by 
the adjacent helices and strands.
Furthermore, far away from the soliton center we  have in analogy with (\ref{asymp})
\[
\kappa_i \ \to \left\{ \begin{matrix}  \mu_1 \  & \mod (2\pi) \ \ \ \ i > s \\  \mu_2 \  & \mod(2\pi) \ \ \ \ i < s
\end{matrix} \right.
\]
%The parameter values are listed in Table I. 
The corresponding  torsion angles are evaluated in terms of the bond angles
using equation (\ref{tauk}).  

Note that in the case of proteins,  the  profile of equation (\ref{eq:bond2}) becomes monotonically increasing 
when we add  multiples of $2\pi$ to the experimental
values. Since the values of  $\kappa_i $'s are defined $\mod(2\pi)$, this does not affect the backbone 
geometry. The integer number of times the monotonically increasing variable $\kappa_i$ 
covers its fundamental domain $[-\pi, \pi)$  counts the number of solitons along the backbone. 
Recall that negative values of $\kappa_i$ are related to positive values of $\kappa_i$ 
by $\mathbb Z_2$ symmetry (\ref{eq:dsgau}).
Finally,  {\it only}  the parameters $\sigma_1$ and  $\sigma_2$ in (\ref{eq:bond2}) are 
intrinsically specific parameters for a given loop. But they specify only the length of the loop, 
not its shape which is determined {\it entirely} by the functional form of equation (\ref{eq:bond2}) 
and, as in the case of $\mu_{1}$ and $\mu_{2}$, 
they are combinations of the parameters in equation (\ref{U}).

In the expression  (\ref{tauk}) of the torsion angles  $\tau_i,\ i=1,2,...,N-3$, 
there are only two independent parameters $u$ and $v$. 
Consequently the profile of $\tau_i$ is determined entirely by $\kappa_i$, and by the structure of the adjacent
regular secondary structures.

It has been shown \cite{Krokhotin-2011}  that most crystallographic protein structures in PDB can be described with
very high precision in terms of such soliton profiles  as their modular building blocks.  Moreover, it has 
been found that in the ensuing soliton profiles, the number of parameters is generically much smaller
 than the number of residues. Thus, the energy function (\ref{E1old}) has a very high predictive power, 
 in describing folded proteins structures in PDB. Its predictions can be subjected to stringent experimental scrutiny,
 both in the case of static and dynamic proteins. 
 
% \vskip 0.2cm

\section{Results}

 We now proceed to demonstrate, that all the concepts and structures we have 
 identified are observed during an
 all-atom simulation of protein folding.
We start with individual atom level scrutiny, even though our goal is to identify and model
those {\it collective} conformational deformations that cause a protein to fold.  We
inquire how does self-organisation, in the case of a protein, 
relate to universal concepts such as
formation of domain walls along spin chains. We study how accurately can the
dynamics and structure of the important collective deformations 
be modelled by soliton profiles such as the one described by the DNLS equation.

We have subjected the single C-chain subunit of the core structure of gp41 \cite{Chan-1997} with PDB code 1AIK and 
residues 628-661,  to detailed all-atom  simulations. We have used the 
GROMACS 4.6.3. package \cite{Hess-2008} with three different force fields --- GROMOS53a6, CHARMM27 and OPLS/AA --- having thus the protein described by
383, 573 and 609 atoms, respectively.  %Accordingly, deformations over 
%short domains can be reliably modeled without any need for special fitting arrangements.
The final production simulation models 80 ns of the protein evolution. We have concluded this to be
sufficient, to identify and analyse the important structural deformations that can take place.

\subsection{Generalities}

\subsubsection{Backbone}

In Figure \ref{fig-6}(a) we show the secondary structure of the final conformations that we have obtained using the three
force fields, in our 80 ns simulations. We observe a clear deformation, in a segment that
consists of the first 10 residues from the N-terminal (upper part in the figure). 
%
%
%                                                                      FIGURE - 6 
%
%
\begin{figure}
\centering
{\begin{minipage}[h]{.45\textwidth}
  %\raggedright
  \includegraphics[trim = 0mm 5mm 20mm -10mm, width=0.9\textwidth,  angle=0]{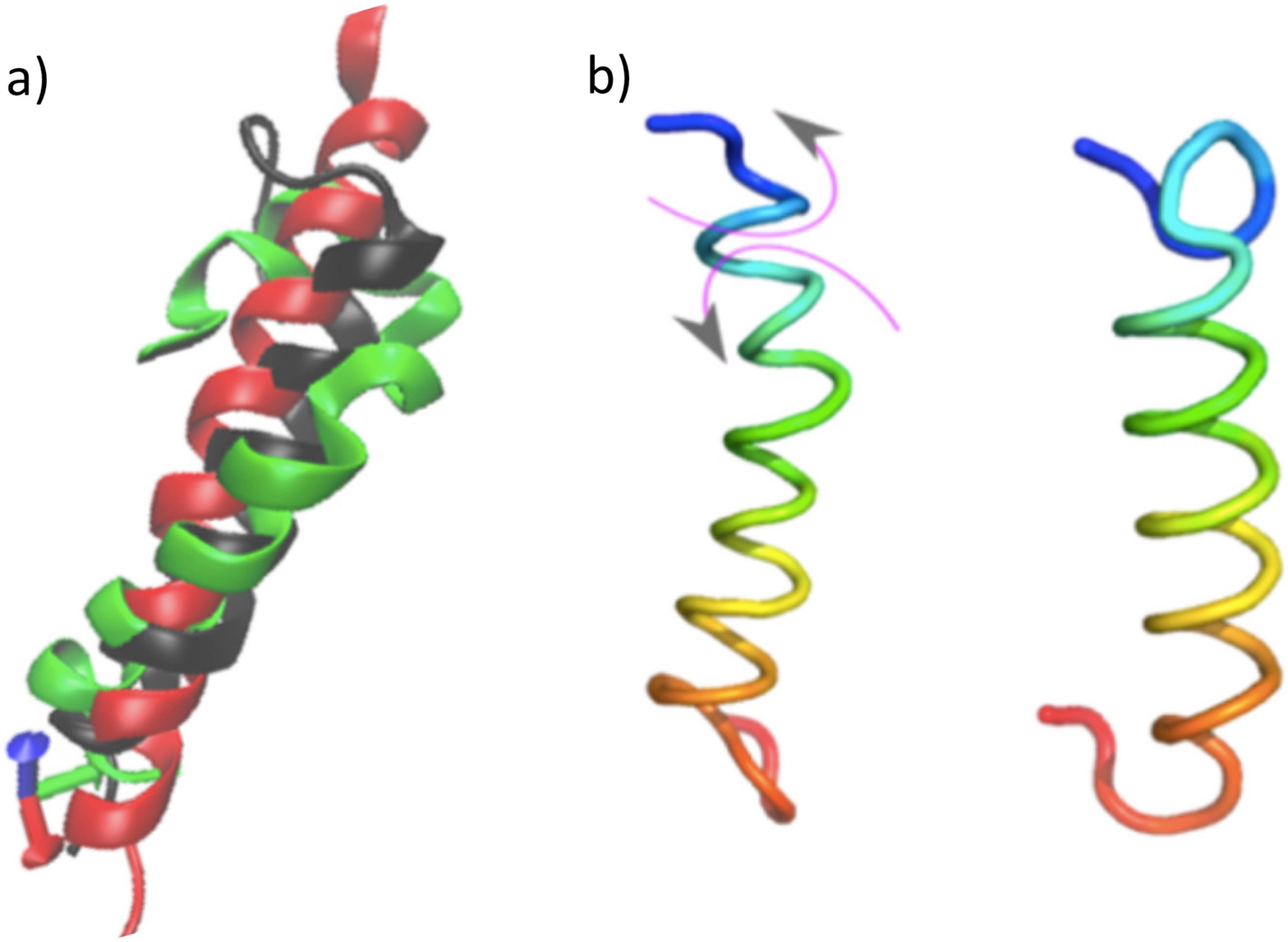}
  \caption{a) Final conformations after  80 ns MD simulations
  with: GROMOS53a6 force field (black);  CHARMM27 force field (red), and OPLS/AA force 
  field (green). b) During the process that we simulate,
  we commonly observe that the N-terminal rotates anticlockwise 
  while the rest of the protein rotates clockwise.}
  \la{fig-6}
\end{minipage}}
\end{figure}

In Figure \ref{fig-7} we display the weighted root-mean-square-deviation (RMSD) of the protein backbone in the three force fields,
%We define this quantity as follows,
\begin{equation}
{\rm RMSD}(t_1, t_2) \ = \ \left[ \frac{1}{M} \sum\limits_{i=1}^n m_i || \mathbf r_i(t_1) - \mathbf r_2(t_2) ||^2 \right]^{1/2}
\la{rmsd}
\end{equation}
%
%
%                                                                      FIGURE - 7 
%
%
\begin{figure}
\centering
{\begin{minipage}[h]{.45\textwidth}
  \raggedright
  \includegraphics[trim = 0mm 5mm 20mm -10mm, width=0.9\textwidth,  angle=0]{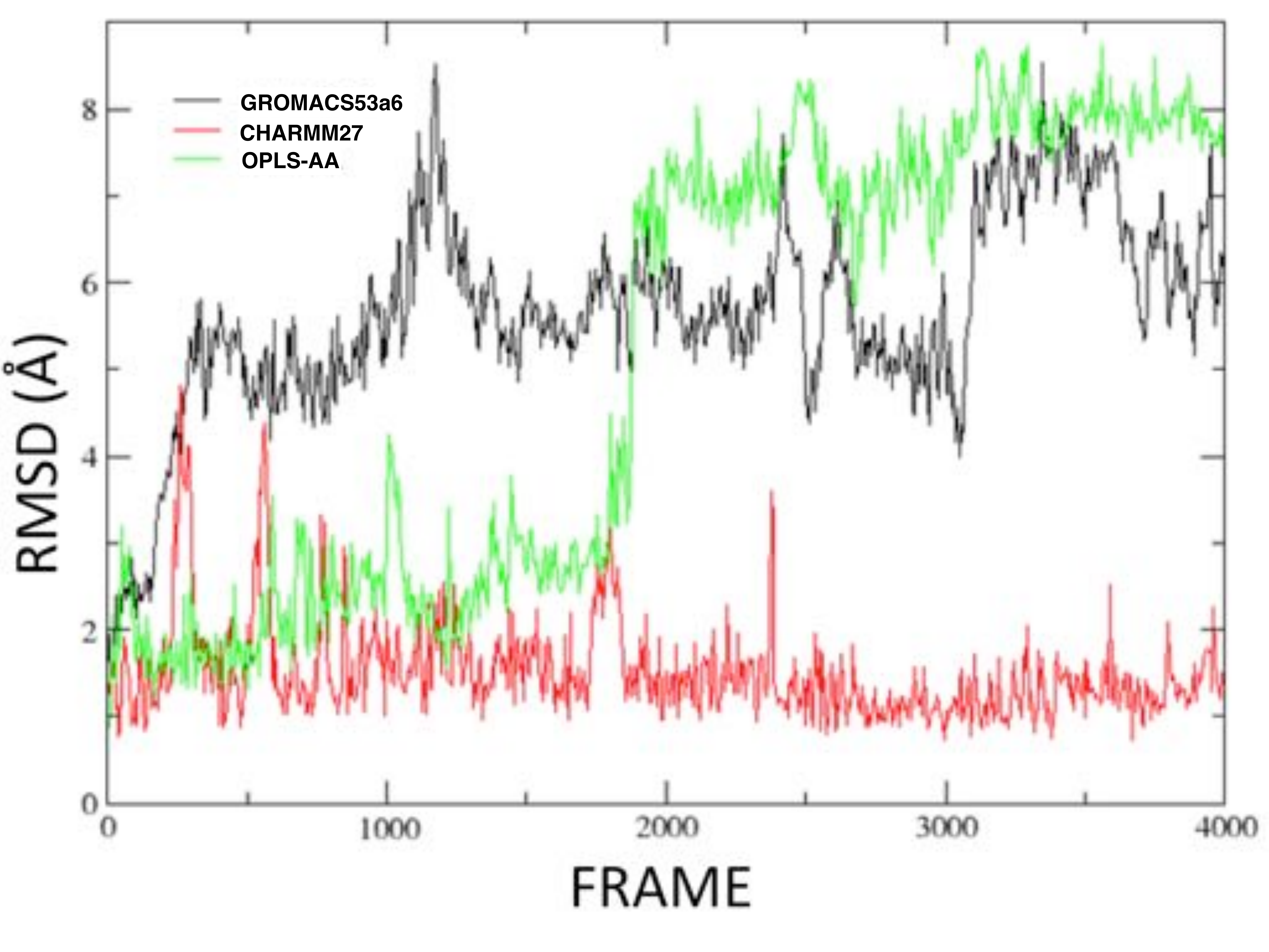}
  \caption{RMSD of the backbone atoms for the
  three force fields: GROMOS53a6 (black), OPLS/AA (green), and CHARMM27 (red).}
  \la{fig-7}
\end{minipage}}
\end{figure}
Here  $\mathbf r_i(t)$ is the position of the atom $i$ at time $t$, and 
\[
M = \sum_{i=1}^n m_i
\]
where $m_i$ are the individual atom masses, and $M$ is the total mass of the backbone.
The deformation is most intense in the GROMOS53a6 force field. With this force field, the initial $\alpha$-helical structure begins collapsing within 4--5 ns.
With the  OPLS/AA force field, we find that the deformation starts after around 40 ns. In the case of
the CHARMM27 force field, the helix tends to remain intact within the selected time range. The deformation begins only after a substantially longer simulation.
Apparently, this force field has a
tendency to produce structures that have an overly $\alpha$-helical content. 
After extended comparisons of the three force fields, including different time steps and 
simulation lengths, we have chosen a 80 ns GROMOS53a6 force field trajectory with  2 fs  time step, 
for the final production simulation that we analyse here. Qualitatively, the results that we present 
are independent of the force field and time step that we have chosen.

\subsubsection{Qualitative considerations}

In figure \ref{fig-8} we show the results from a {\tt do.dssp}  \cite{Hess-2008} secondary structure analysis, in the case of the  
GROMOS53a6 simulation. 
%
%
%                                                                  FIGURE 8
%
%
\begin{figure}
\centering
{\begin{minipage}[h]{.45\textwidth}
  \raggedright
  \includegraphics[trim = 0mm 5mm 20mm -10mm, width=0.9\textwidth,  angle=0]{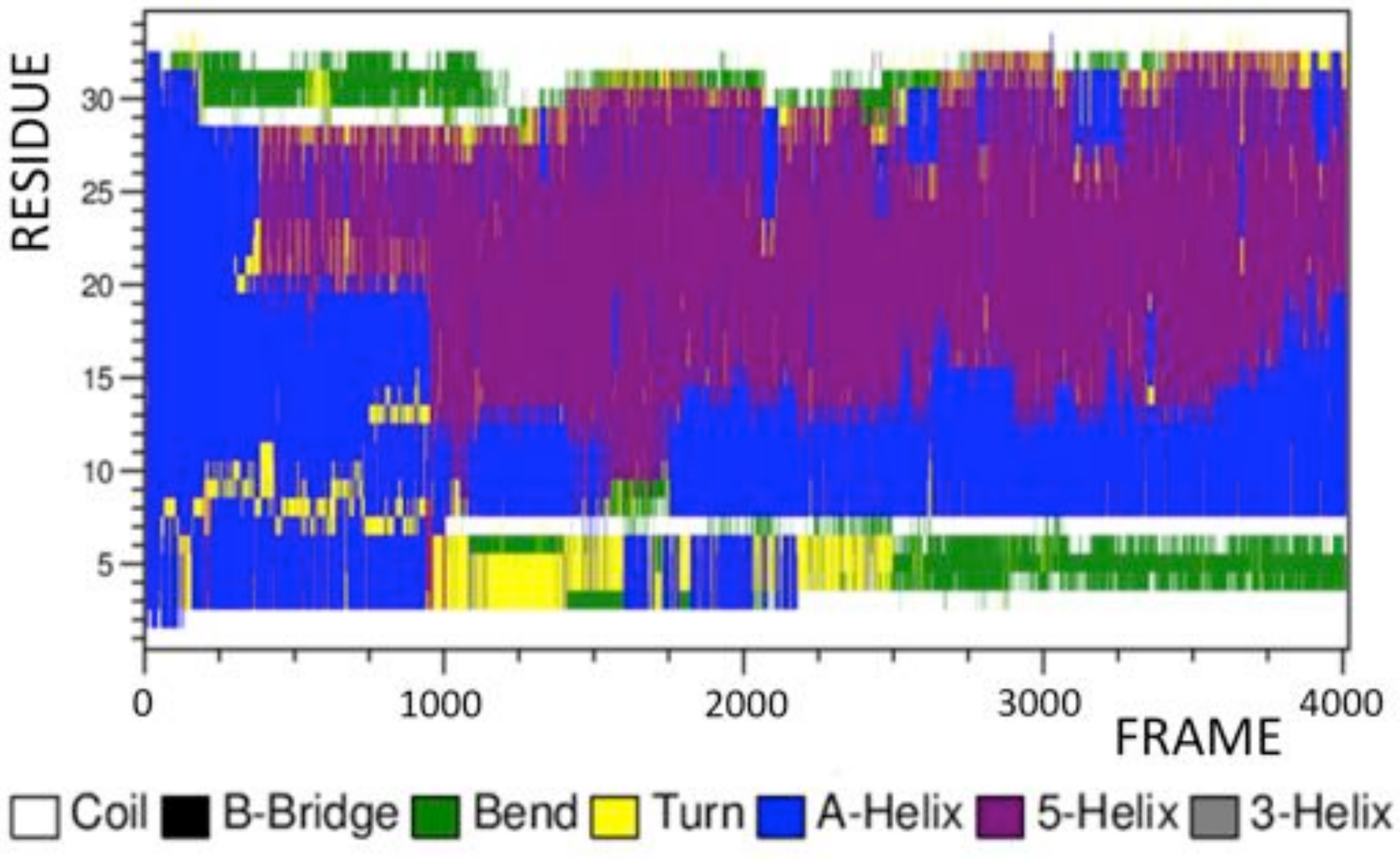}
  \caption{Secondary structure analysis using {\tt do.dssp} along a 80 ns trajectory, produced using the 
  GROMOS53a6 force field. The PDB residues 628-661 are labeled 0-33.}
  \la{fig-8}
\end{minipage}}
\end{figure}
The following {\it qualitative} observations can be made:

$\bullet$ After around 4-5 ns corresponding to frames 200-250, there is an initial formation of a coil structure,
according to  {\tt do.dssp} 
classification. The coil 
becomes initially stabilised around residue number 29 which corresponds to amino acid number 656
in the PDB file.  The coil is connected to the C-terminal with a bend. At 
around 8 ns (around frame 400) there are helical fluctuations in this structure towards N-terminal,  
and  at around 22-24 ns (frames 1100-1200) 
the coil structure moves back towards the C-terminal. The motion takes place in two steps, at around frame 1200 and
then again at around frame 2800 after which the coil disappears, by merging into 
the apparently random fluctuations of the C-terminal residues. 

At the level of {\tt do.dssp} secondary structure analysis the coil which emerges near
the C-terminal and propagates along the backbone, 
is putatively akin the propagating loop structure that has been previously 
identified and studied in coarse grained UNRES simulations of the protein G related albumin-binding domain 
with PDB code 1GAB \cite{Krokhotin-2014-a,Sieradzan-2014}. In particular the UNRES simulation \cite{Krokhotin-2014-a} identifies
a displaced protein loop as a localised structure with a profile that can be described
by the soliton solution of the discrete nonlinear Schr\"odinger equation (\ref{tauk}), (\ref{nlse}). 
The simulation demonstrated that when the loop-soliton moves along the protein lattice, with cells
matching the residues, there are waves  that are emitted in its  wake 
as vibrations in the lattice structure. These 
waves drain the kinetic energy of the soliton, and cause 
it to decelerate. Eventually the kinetic energy of the soliton becomes depleted, and it can no longer cross over the 
energy barriers between lattice cells and becomes localised around a particular set of lattice cells. The energy barriers that
prevent the soliton from translating along the backbone lattice were identified as Peierls-Nabarro barriers 
\cite{Peierls-1940,Nabarro-1947,Nabarro-1997}  in \cite{Sieradzan-2014}.  

In the present case of the C-terminal coil, there is apparently
a Peierls-Nabarro barrier that stops and prevents the coil that is supposedly modelled by a DNLS,
from  propagating away from the C-terminal beyond the residues 28-29.
Instead it becomes initially trapped, then moves towards the C-terminal and dissolves there.  The soliton moves step-wise,
it's crossing-over the ensuing Peierls-Nabarro barriers is boosted by thermal fluctuations. The soliton 
crosses a barrier whenever the amplitude of its thermal fluctuations exceeds the barrier-specific threshold value.

$\bullet$ At around the same time when the C-terminal coil forms, we observe a turn 
deformation that forms and proceeds away from the N-terminal, and then fluctuates thermally between 
residues 5 and 10. After around 20 ns (frame 1000) of simulation time, 
there is a rapid extended turn-like fluctuation that connects the N-terminal with a localised structure 
which is identified as a short coil by  {\tt do.dssp}. This is an apparent DNLS soliton, 
emerging at the end of the extended turn-like structure and  stabilising 
around the residues 6-8 (residues 634-636 in the PDB file).
We observe initially relatively strong fluctuations in the residues between the putative soliton and the N-terminal.
But the amplitudes of these fluctuations become damped and after around 50 ns (frame 2500) 
there are only minor fluctuations in the soliton. There is a bend between the soliton and the N-terminus which
constitutes a Peierls-Nabarro barrier, high enough to prevent 
the soliton  from moving towards the N-terminal, step-wise by thermal fluctuations.

\subsubsection{Backbone folding index}

We proceed to analyse the dynamics {\it quantitatively}, and we start with
the backbone folding index (\ref{eq:Gamma}).  For this 
we have 
divided the backbone into segments of varying length, and computed the folding index over the segments during 
the 80 ns time evolution. Examples of results  are shown in Figure \ref{fig-9}, where  we plot the 
numerical values of the folding index density (\ref{eq:foldind}). 
%
%
%                                                       figure-9b
%
%
\begin{figure}
\centering
{\begin{minipage}[h]{.45\textwidth}
  \raggedright
  \includegraphics[trim = 0mm 5mm 20mm -10mm, width=0.9\textwidth,  angle=0]{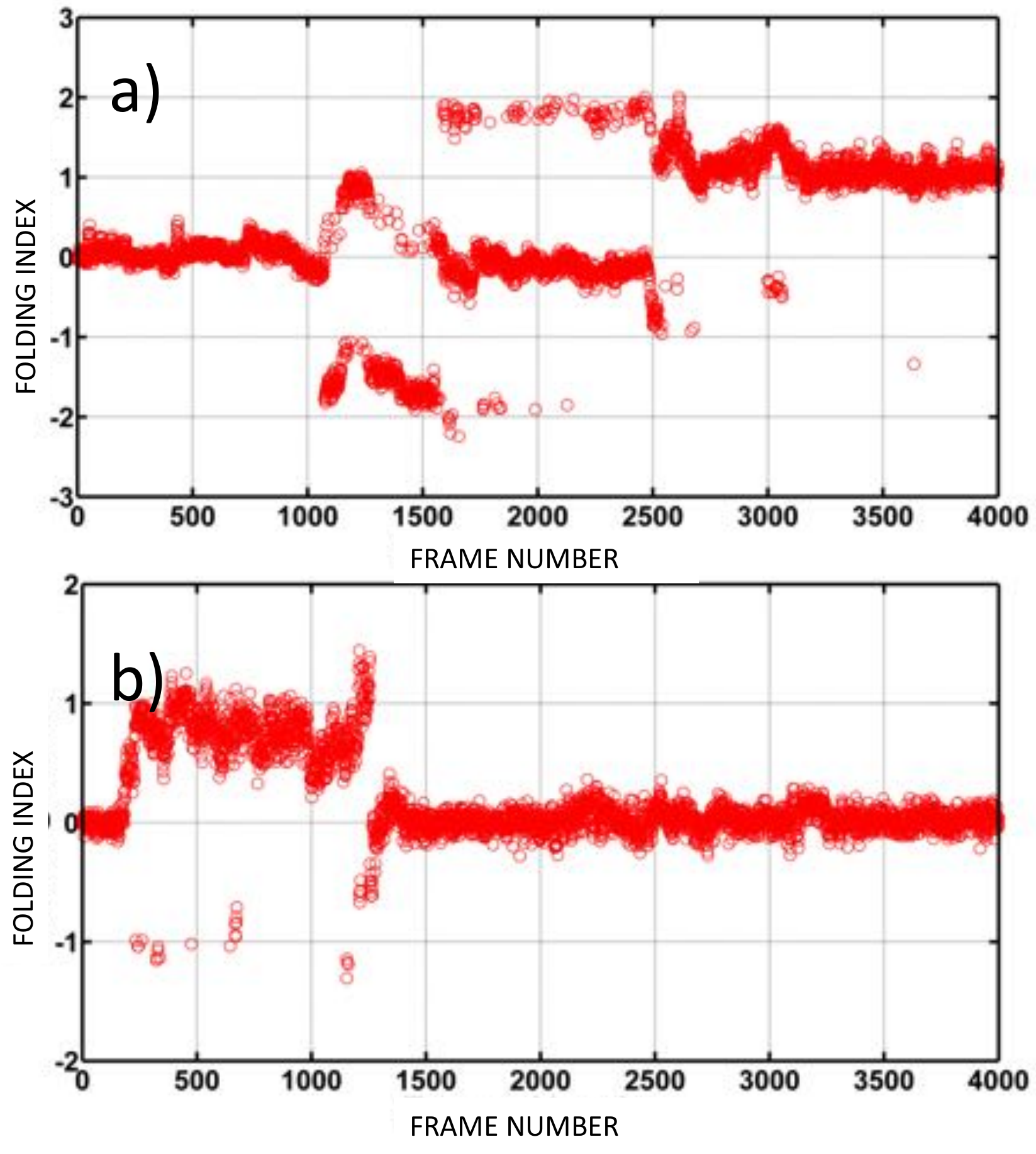}
  \caption{The folding index density (\ref{eq:foldind}) evaluated over two segments of the 1AIK backbone: (a) the segment 4-11 (residues 632-639 in PDB), and (b)  the segment  
  24-30 (residues 652-658 in PDB). }
  \la{fig-9}
\end{minipage}}
\end{figure}
The first  segment consists of the sites 4-11 corresponding to residues 632-639 in PDB.
This segment covers the N-terminal soliton structure (see Figure \ref{fig-8}).  The second segment consists of the
sites 24-30 (652-658 in PDB).  This covers the segment where the C-terminal coil initially appears 
in Figure \ref{fig-8}. We observe the following:

{\bf Figure \ref{fig-9}(a): } Initially, the folding index of the segment 4-11 vanishes. But in the vicinity of frame 1000,  
coinciding with the formation of the N-terminal
coil/soliton in figure \ref{fig-8}, the folding index starts fluctuating between the values $Ind_f = \pm 1$ and 
$Ind_f = -2$. We note how the pattern of the oscillations reflects the structural changes in the region 
between the coil and the N-terminal, shown in figure \ref{fig-8}: 
There are first fluctuations between turn and bend, during frames 1000-1500. 
When the oscillations in the values of the  folding index diminish and vanish near the frame 1500, 
we observe a formation of helical structure between the coil and the N-terminal in Figure \ref{fig-8}.
The folding index then starts oscillating between the values $Ind_f = 0, 2$, and this corresponds to a
frame segment where the helix converts into a turn in the Figure \ref{fig-8}. At around the frame 2500, the folding index
finally stabilises to the final value $Ind_f =  +1$. This stabilisation concurs with the formation of a bend 
between the coil and the N-terminal, in Figure \ref{fig-8}.

{\bf Figure \ref{fig-9}(b):} We observe the increase of folding index from $Ind_f = 0$ to  $Ind_f = +1$ near frame 200, and subsequent
decrease back to $Ind_f = 0$ near frame 1300. According to figure \ref{fig-8}, 
these transitions  coincide with the appearance of the C-terminal soliton, 
and its subsequent propagation towards the C-terminal, away from the segment 24-30.

Accordingly, we have found that the variations in the values of the folding index, in particular in 
the case of the N-terminal soliton, coincide with the structural deformations that take place along the 
backbone segment which is located between the soliton and the N-terminal.   In particular, 
the final stabilisation of the folding index concurs with the crossing over the Peierls-Nabarro barrier and subsequent 
stabilisation of the soliton, according to figure \ref{fig-8}. Moreover, 
the C-terminal soliton emerges with $Ind_f = +1$ and remains stable until the Peierls-Nabarro barrier
crossing takes place. The evolution of the ensuing  folding index is also fully in line with 
the results that we deduce from {\tt do.dssp}.  We conclude that the behaviour in the backbone segment,  surrounding
the soliton, directly correlates with the topological character of the soliton, in both cases.

Finally, in Figure \ref{fig-10} we show the folding index density (\ref{eq:foldind}) over the {\it entire} backbone and for all the 
4000 frames.
%
%
%                                                       figure-10
%
%
\begin{figure}
\centering
{\begin{minipage}[h]{.45\textwidth}
  \raggedright
  \includegraphics[trim = 0mm 5mm 20mm -10mm, width=0.9\textwidth,  angle=0]{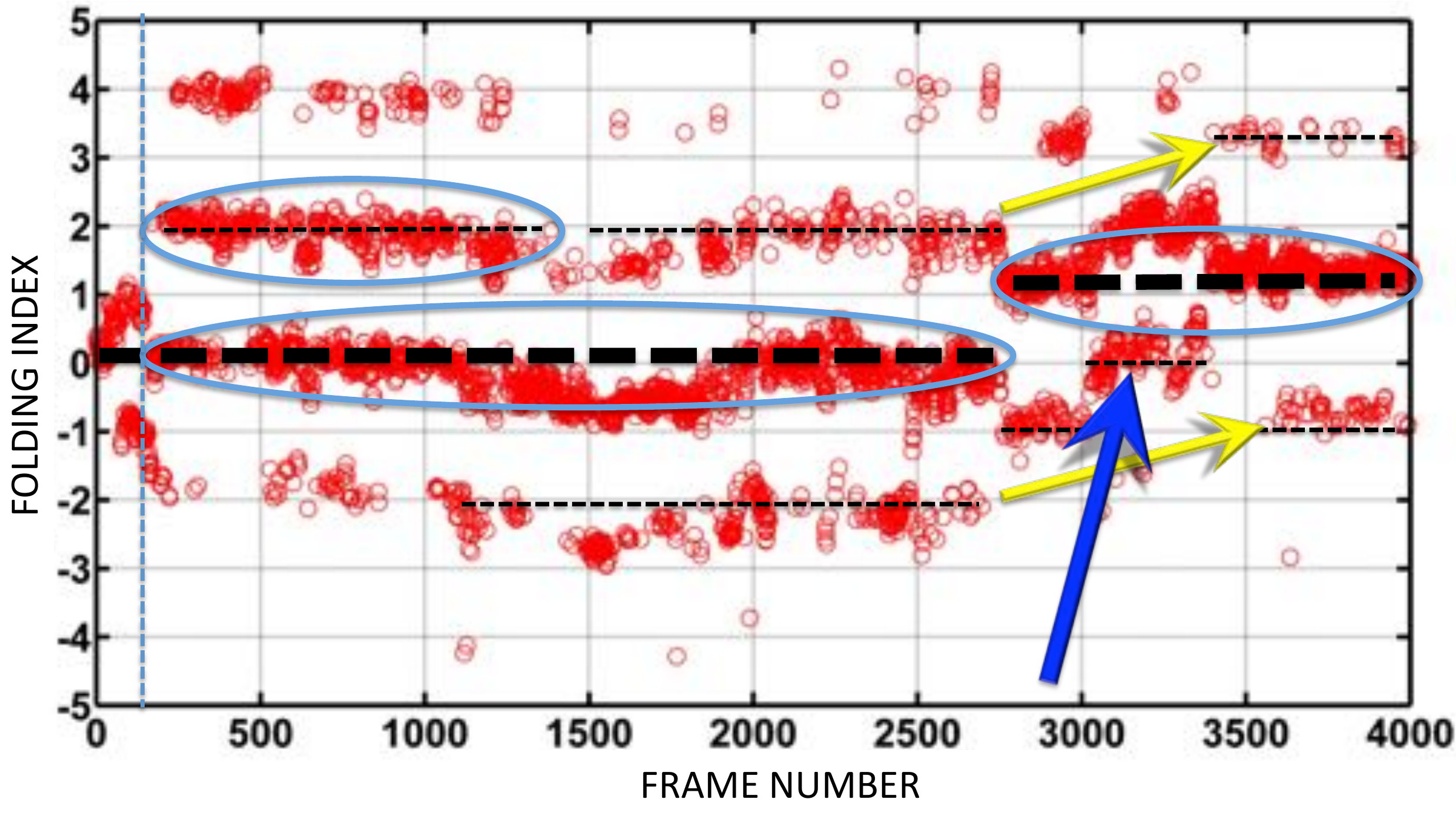}
  \caption{The folding index density (\ref{eq:foldind}) evaluated over the entire backbone for all 4000 frames. Some of the
  major features have been high-lighted.}
  \la{fig-10}
\end{minipage}}
\end{figure}
We observe that:

$\bullet$ Initially, the folding index vanishes. This is
consistent with the $\alpha$-helical structure of the 1AIK sub-chain. But there is a sudden
initial transition to the value $Ind_f = +1$, presumably reflecting the initial stages of C-terminal soliton formation. 

$\bullet$ Up until the frame $\sim 2700$ the folding index tends to vanish. But there are fluctuations, mainly between values
$\pm 2$  which reflect the 
various processed that take place near the terminals. 

$\bullet$ In the vicinity of frame 2700 there is a transition, and the value of the folding index  starts stabilising toward the value
$Ind_f = +1$. This stabilisation concurs with the stabilisation of the N-terminal soliton, and the final departure  of the C-terminal
soliton. The fluctuations also shift, oscillating between $Ind_f=+3$ and $Ind_f=+1$ and this 
shift is identified  by the yellow arrows in the figure.

In Figure \ref{fig-11} we show a close-up to the last 50 frames in figure \ref{fig-10}. It confirms 
the stabilisation of the folding index towards the value $Ind_f = +1$, with occasional fluctuations where $Ind_f = +3$ or 
$Ind_f = -1$.
%
%
%                                                       figure-11
%
%
\begin{figure}
\centering
{\begin{minipage}[h]{.45\textwidth}
  \raggedright
  \includegraphics[trim = 0mm 5mm 20mm -10mm, width=0.9\textwidth,  angle=0]{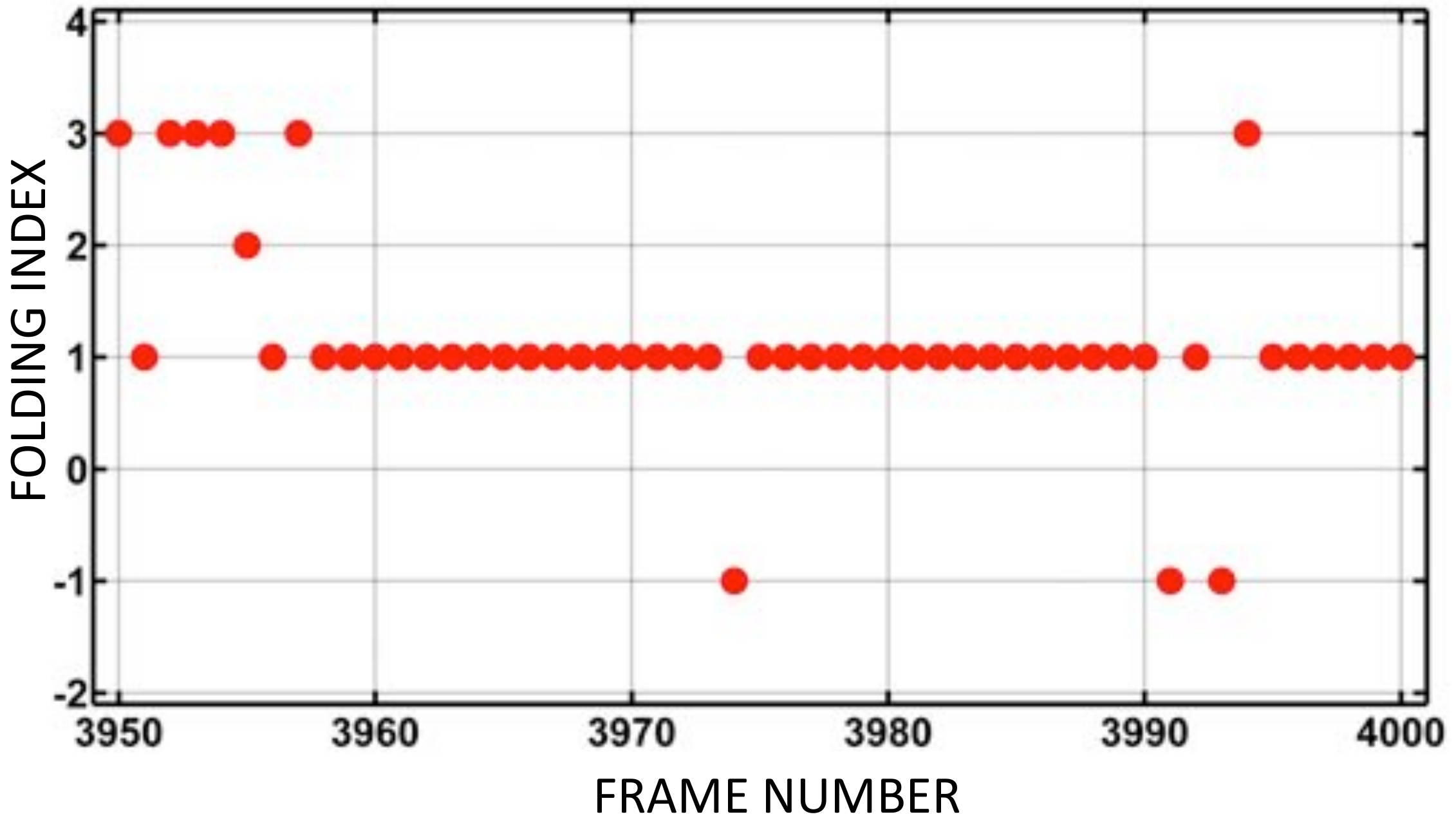}
  \caption{The close-up of segments 3050-4000 in the folding index density of figure \ref{fig-10}. }
  \la{fig-11}
\end{minipage}}
\end{figure}
In Figure \ref{fig-12},
%
%
%                                                       figure-12
%
%
\begin{figure}
\centering
{\begin{minipage}[h]{.45\textwidth}
  %\raggedright
  \includegraphics[trim = 0mm 5mm 20mm -10mm, width=0.9\textwidth,  angle=0]{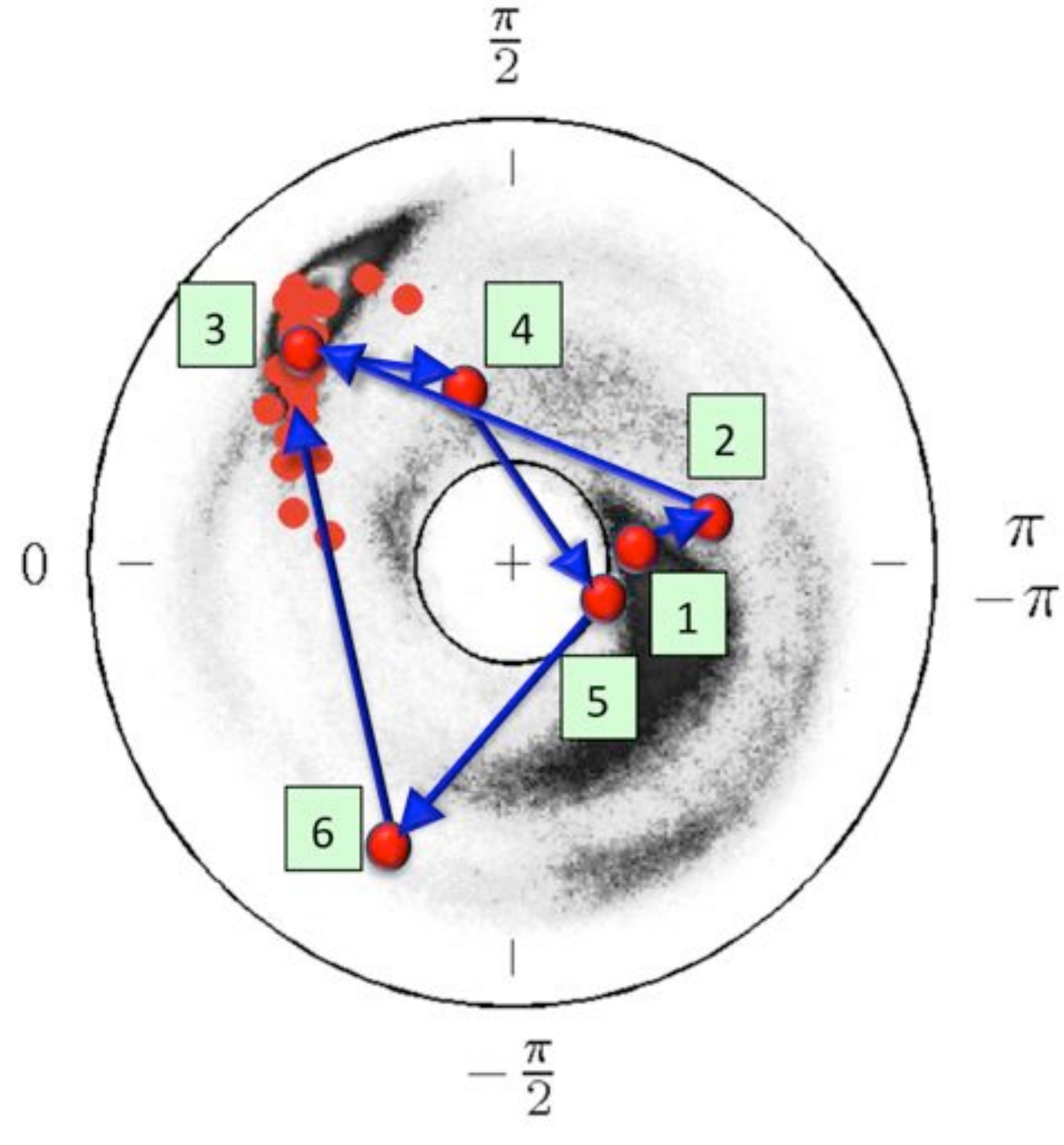}
  \caption{The trajectory of the frame 4000 on the ($\kappa,\tau$) landscape of figure \ref{fig-2}, following figure \ref{fig-3} a). Note
  that after residue 6, the remaining residues are all located closely, in the $\alpha$-helical region.}
  \la{fig-12}
\end{minipage}}
\end{figure}
following Figure \ref{fig-3}(a), we show the full trajectory for the entire final frame 4000. The trajectory starts from the N-terminal
which is located in the $\beta$-stranded region of Figure \ref{fig-2}. It moves over to the $\alpha$-helical region, then return to the 
$\beta$-stranded region to encircle the north pole. Finally, the trajectory  merges and ends with the $\alpha$-helical region.
The trajectory  confirms that the final structure at frame 4000 indeed does support a twisting $\Delta \tau = +\pi$ and that 
the twisting is furthermore located at the N-terminal soliton.

The stabilisation of the folding index to the value $Ind_f = +1$ confirms the global character of the remaining N-teminal soliton structure:
There is a total twisting by $\Delta \tau = +\pi$ along the final backbone, in comparison to the initial configuration and including
the terminal residues,  and this twisting is localised on the N-terminal soliton. Moreover, we observe that the N-terminal residues
are in a $\beta$-stranded position while the C-terminal residues are in the $\alpha$-helical position.

\subsubsection{Side chain analysis}

Figure \ref{fig-4} shows the landscape of the ground state (crystallographic structure)  C$\beta$ atom directions 
in the C$\alpha$ centered discrete Frenet frames. Figure \ref{fig-13} shows how the directions of the C$\beta$ evolve  
during our entire GROMOS53a6 simulation.  
%
%
%
%                                                      figure-13
%
%
\begin{figure}
\centering
{\begin{minipage}[h]{.45\textwidth}
  %\raggedright
  \includegraphics[trim = 0mm 5mm 20mm -10mm, width=0.9\textwidth,  angle=0]{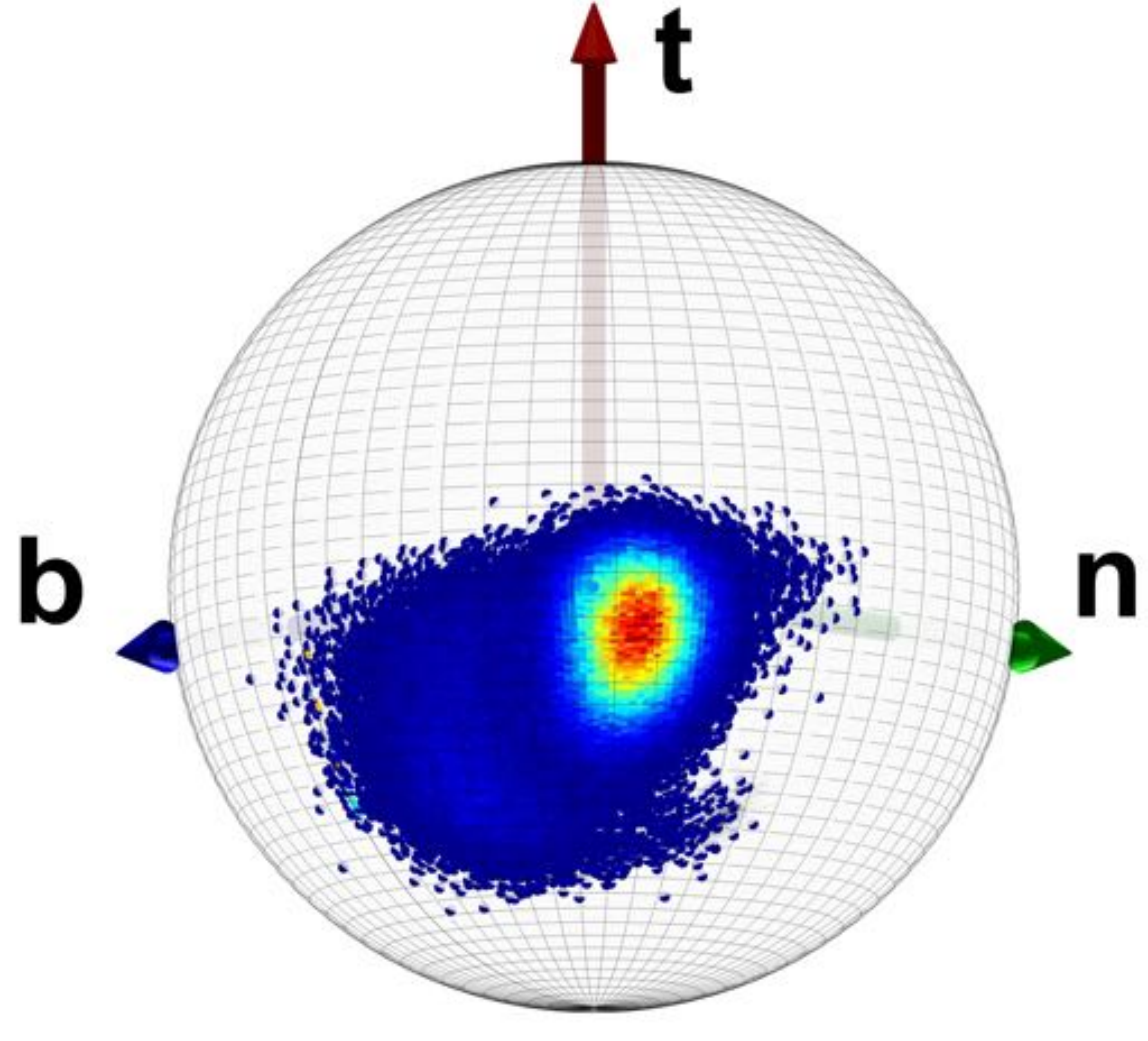}
  \caption{The dynamical landscape of all the C$\beta$ atoms during the entire 80 ns GROMOS53a6
  simulation. A comparison with Figure \ref{fig-4} establishes the presence of strong correlations between the 
  backbone C$\alpha$ and the side chain C$\beta$  geometries during the entire process.}
  \la{fig-13}
\end{minipage}}
\end{figure}

We find it remarkable how similar the dynamical landscape of Figure \ref{fig-13} is with the static ground state
landscape shown in Figure \ref{fig-4}: The direction of C$\beta$ nutates {\it tightly} around its static ground state landscape.  
Clearly, there must be strong correlations  between the
backbone C$\alpha$ and side chain C$\beta$,  during the entire dynamics.
Accordingly, the information content in the angles $\eta_i$ in (\ref{etai}), (\ref{etai2}) should correlate
strongly with the C$\alpha$ geometry changes during the dynamical process.
  
More generally, we expect that the various backbone and side chain spin models that we have introduced, 
are all in the same dynamical universality class.

Figure \ref{fig-14} 
%
%
%                                                                        FIGURE 14
%
%
%
\begin{figure}
\centering
{\begin{minipage}[h]{.45\textwidth}
  \raggedright
  \includegraphics[trim = 0mm 5mm 20mm -10mm, width=0.95\textwidth,  angle=0]{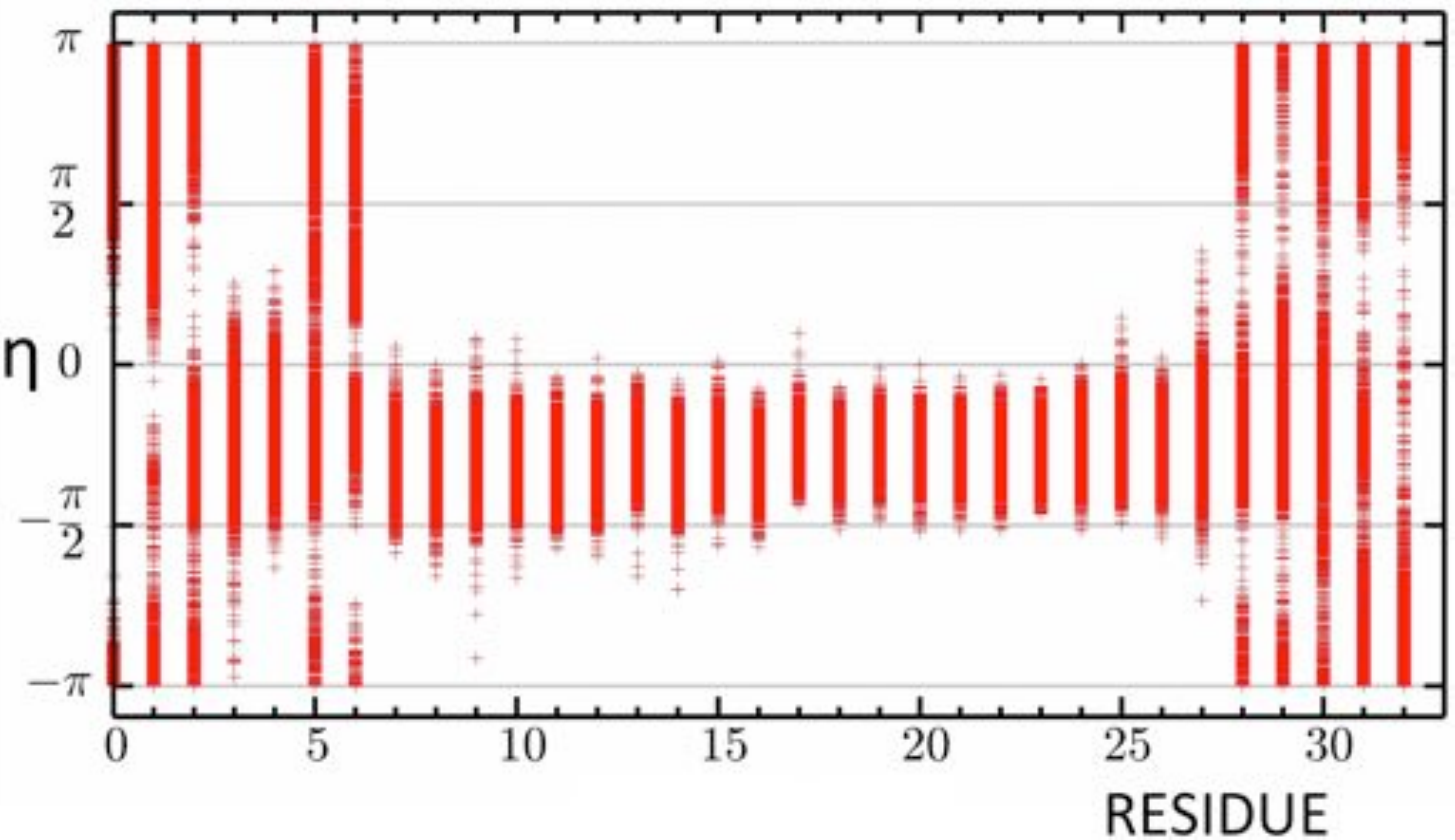}
  \caption{The residue-wise distribution of the angles $\eta_i$ in (\ref{etai}), during the entire GROMOS53a6 80 
  ns run. The horizontal axis labels the residues and the vertical axis is the value of the angle $\eta$
  in radians.}
  \la{fig-14}
\end{minipage}}
\end{figure}
shows the residue-wise accumulated distribution of all the individual angles $\eta_i$ in (\ref{etai}) {\it i.e.} the ensuing landscape
of the individual $\eta_i$ during the entire 80 ns 
GROMOS53a6 simulation.  The conclusions that can be deduced from this
figure are in line with those in Figure \ref{fig-8}. In particular,  we 
observe the presence of the  N-terminal soliton, how it is centered around residues 5-6. We 
also observe that the residues between sites 6-27 are in a helical position during the
entire time evolution.  We also observe the merging of the C-terminal soliton with the fluctuations of the C-terminal.

Figure \ref{fig-15}  shows the time resolved  landscape of all the $\eta_i$ angles. 
%
%
%                                                                         FIGURE 15
%
%
\begin{figure}
\centering
{\begin{minipage}[h]{.45\textwidth}
  \raggedright
  \includegraphics[trim = 0mm 5mm 20mm -10mm, width=0.9\textwidth,  angle=0]{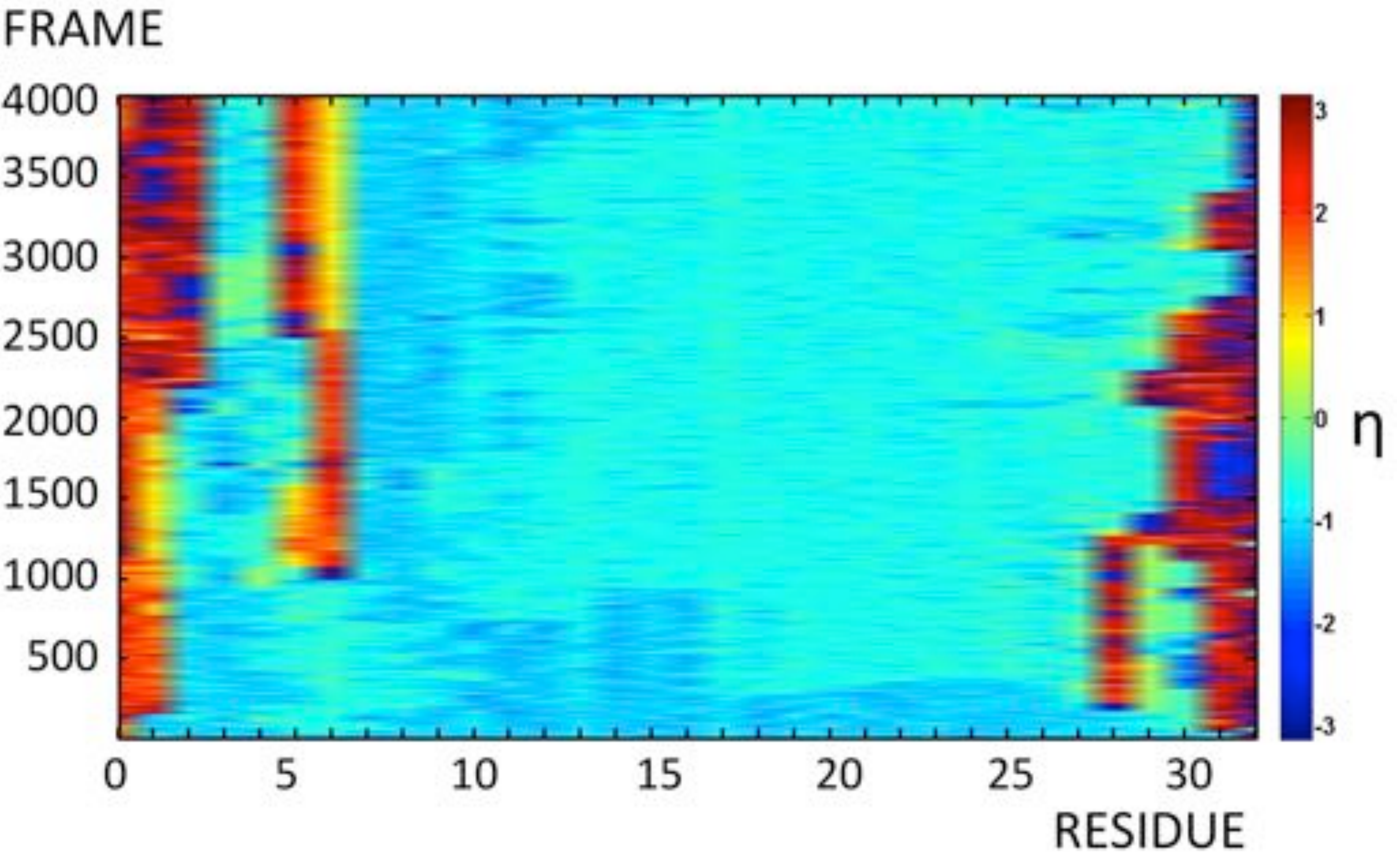}
  \caption{The time resolved evolution of all the individual angles $\eta_i$. Note the similarity with Figure \ref{fig-8}. }
  \la{fig-15}
\end{minipage}}
\end{figure}
There is a remarkable similarity between this figure, and the figure obtained from the
{\tt do.dssp} backbone analysis shown in figure \ref{fig-8}. In particular, the formation and stabilisation
of the  N-terminal soliton around sites 5-6 is clearly visible. The appearance of the
C-terminal soliton and its subsequent evolution is similarly visible: we observe how this soliton 
is formed at around frame 200, and then propagates  towards the C-terminal in a stepwise manner, crossing over
the various Peierls-Nabarro barriers and  eventually merging with the C-terminal thermal fluctuations.

Finally we note  the apparent 
similarities between the structure of the landscape in Figure \ref{fig-15} and the behaviour of the folding index density
in Figure \ref{fig-10}. 

\subsection{Details}

We proceed to analyse  the detailed properties of soliton structure and formation.  Our main focus  will be on the
N-terminal soliton structure. We are particularly interested  in  
the phenomena that take place when the soliton is formed, {\it i.e.} the 
vicinity of the frame $\sim$1000, and when the soliton moves over to a Peierls-Nabarro barrier and stabilises,
{\it i.e.} the  vicinity of the frame $\sim$2500. 

\subsubsection{C-terminal side chain soliton}

We start with a closeup of the C-terminal part in Figure \ref{fig-15}, shown in Figure \ref{fig-16}.
%
%
%                                                                         FIGURE 16
%
%
\begin{figure}
\centering
{
\begin{minipage}[t]{.45\textwidth}
  \raggedright
  \includegraphics[trim = 0mm 5mm 20mm -10mm, width=0.9\textwidth,  angle=0]{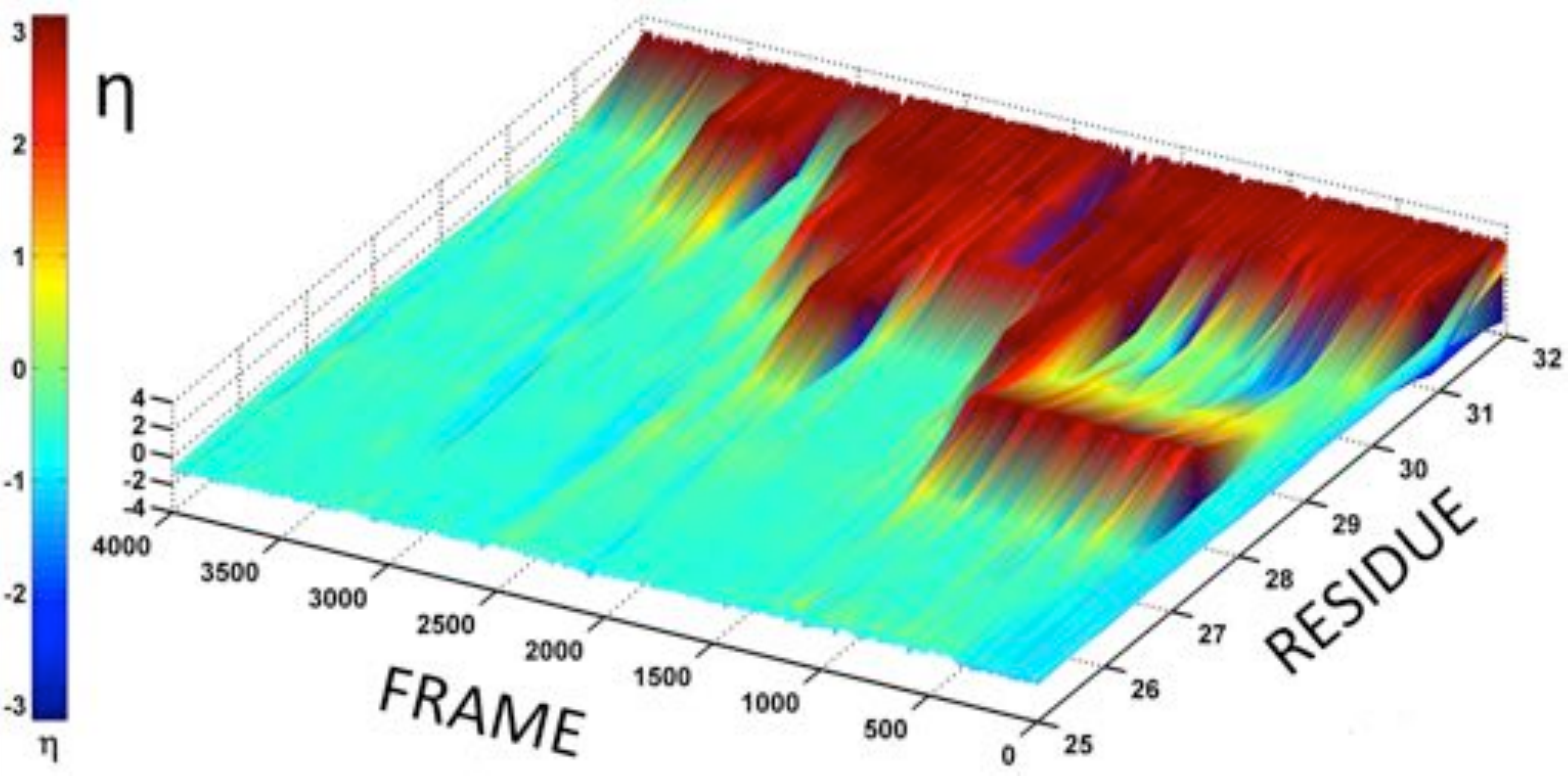}
  \caption{A close-up of the time resolved evolution of the individual angles $\eta_i$, in the case of the C-terminal soliton
  structure; only the last 9 residues along the backbone are shown.   }
  \la{fig-16}
\end{minipage}
}
\end{figure}
We observe how, in terms of the side chain $\eta_i$ angles, the soliton structure which forms with center at  residue
28 subsequently propagates back-and forth,
in a step-wise manner, towards the C-terminal. Eventually it merges with the terminal,  and dissolves into its 
fluctuations. The soliton motion is fully in line with Figure \ref{fig-8}, and the initial motion is consistent with 
the folding index analysis in Figure \ref{fig-9}(b). In particular,   the step-wise
propagation of the soliton is fully consistent and in line with the presence of Peierls-Nabarro barriers. These barriers are
high enough to trap the soliton momentarily, but low enough for the soliton  
to eventually cross over them when its thermal excitation energy fluctuates to high
enough value.  

We remind that the present simulations have been performed at 290 K. We have chosen this relatively low temperature value,
from the {\it in vivo} perspective,  in order to restrain the soliton mobility and to dampen noisy thermal fluctuations.

\subsubsection{N-terminal side chain soliton}

Figure \ref{fig-17} shows a close-up of the N-terminal part in Figure \ref{fig-15}, from two complementary perspectives.
%
%
%                                                                         FIGURE 17
%
%
\begin{figure}
\centering
{
\begin{minipage}[t]{.45\textwidth}
  \raggedright
  \includegraphics[trim = 0mm 5mm 20mm -10mm, width=0.9\textwidth,  angle=0]{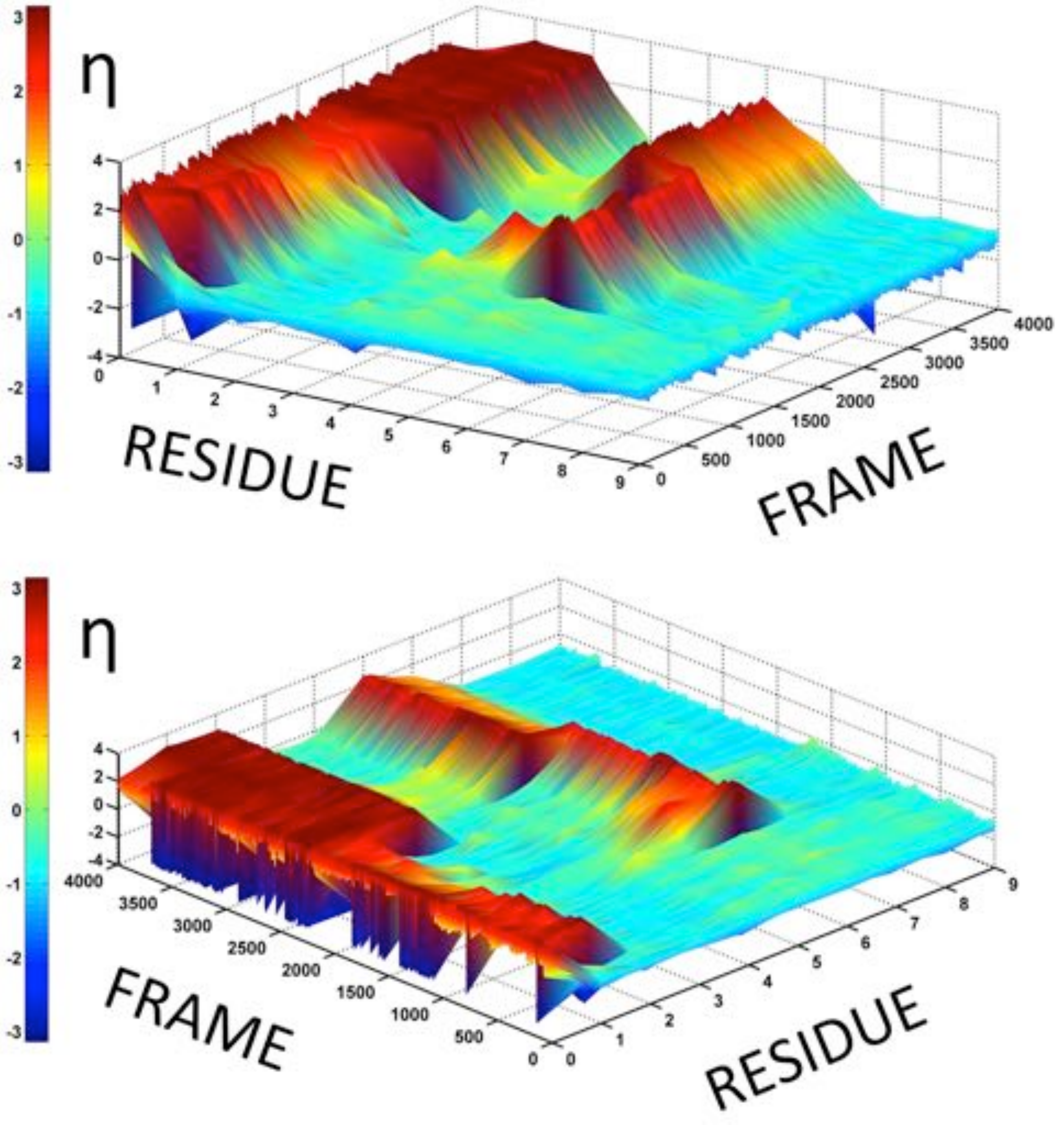}
  \caption{The time resolved evolution of the individual angles $\eta_i$, for the N-terminal soliton structure. The 
  two figures show the same data, but from different perspectives, for the first 10 residues.}
  \la{fig-17}
\end{minipage}
}
\end{figure}
The soliton appears in the vicinity of frame 1000. It
subsequently translates one residue 
towards the N-terminal, in the vicinity of frame 2500. This is an apparent crossing
over a Peierls-Nabarro barrier by thermal fluctuation,  and it is followed by a stabilisation of the soliton  
at the final position. Note  the
correlation between the soliton motion and the extent of the  N-terminal fluctuations.

Figure \ref{fig-18} 
%
%
%                                                                         FIGURE 18
%
%
\begin{figure}
\centering
{\begin{minipage}[h!]{.45\textwidth}
  \raggedright
  \includegraphics[trim = 0mm 5mm 20mm -10mm, width=0.9\textwidth,  angle=0]{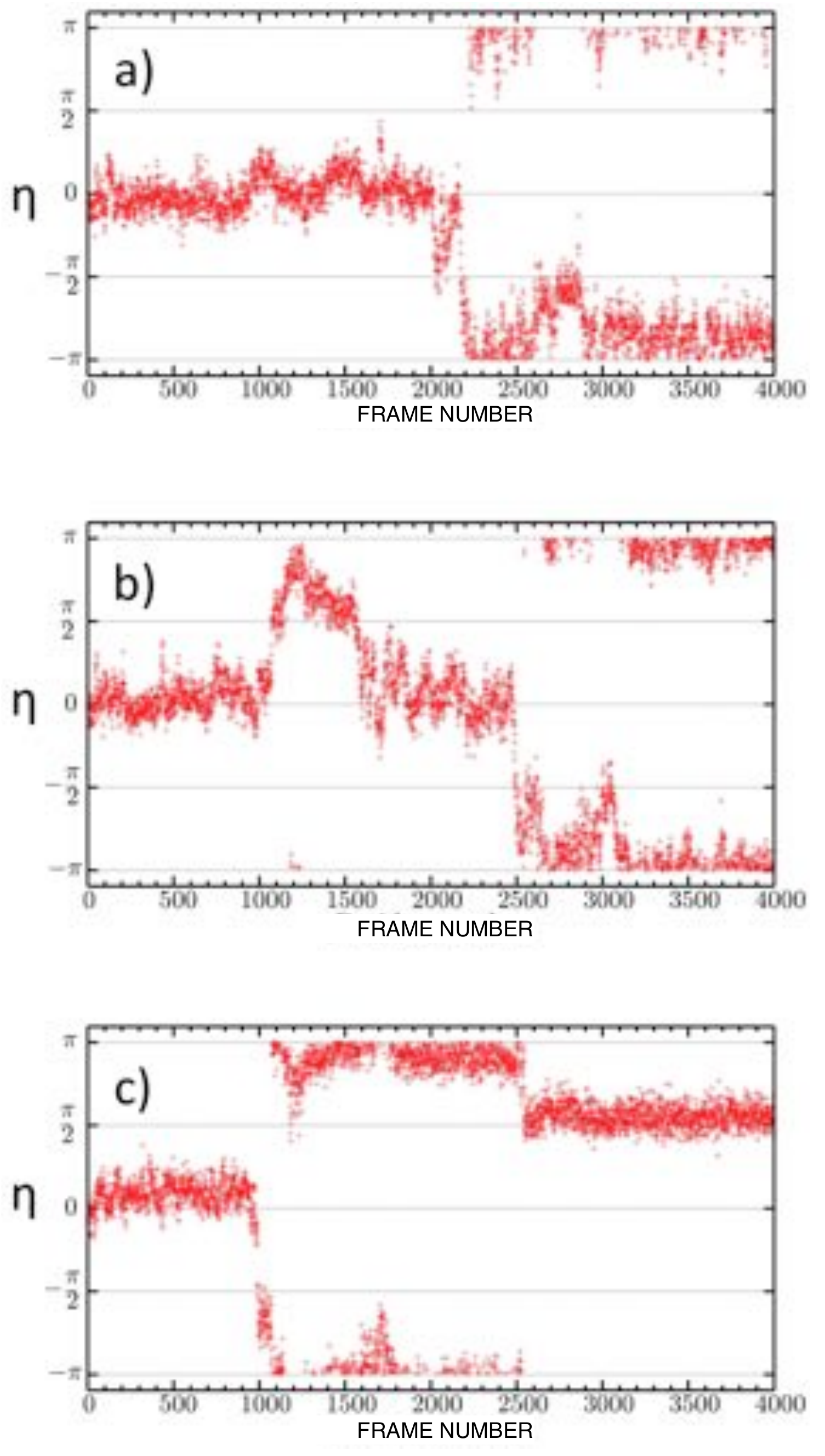}
  \caption{Fluctuations in the values of the $\eta_i$ angle during the entire dynamical process, around
  the average initial value evaluated from the original PDB structure.
  Panels (a)--(c) correspond to residues $i=2,5,6$ respectively. Note that the values are in the range [$-\pi,\pi$] mod($2\pi$).
   }
  \la{fig-18}
\end{minipage}}
\end{figure}
shows the evolution of the individual $\eta_i$ angles in (\ref{eq:etaprime}), in the case of the N-terminal 
soliton structure. The panels 
display the deviation of $\eta_i$ from the initial average value for residues $i=2,5,6$, which we have found to be those of primary interest.  
For $i=3,4$ and for $i=7$ and larger, the deviations from the initial average value fluctuate around zero.

In Figure \ref{fig-19} we scrutinise those segments of Figure \ref{fig-18}, where the $mod(2\pi)$ branch of the angle 
needs to be carefully resolved.
%
%
%                                                                         FIGURE 19
%
%
\begin{figure}
\centering
{\begin{minipage}[h]{.45\textwidth}
  \raggedright
  \includegraphics[trim = 0mm 5mm 20mm -10mm, width=0.9\textwidth,  angle=0]{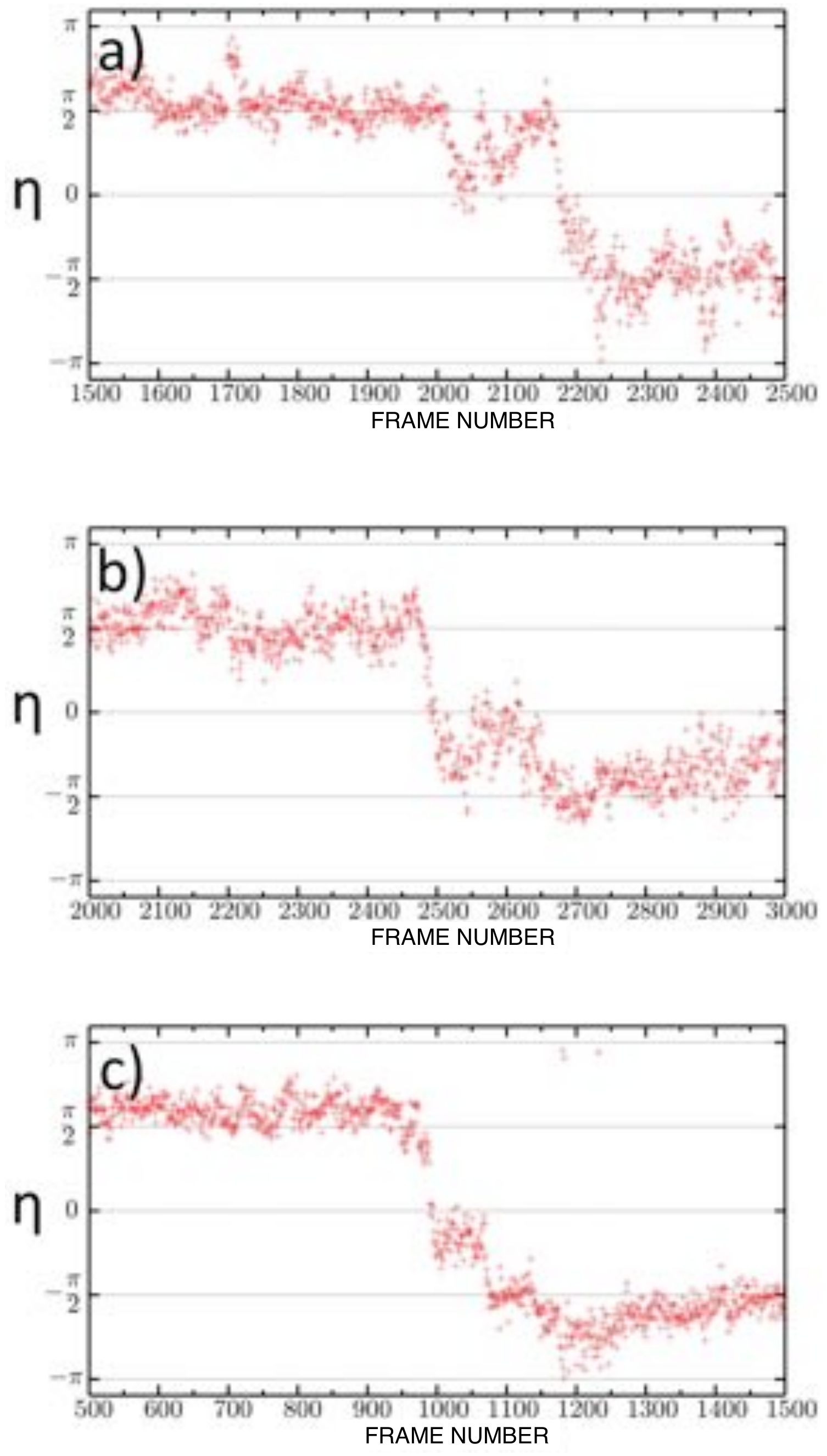}
  \caption{Details of the corresponding panels in Figure \ref{fig-18}, where a careful scrutiny is needed due to the multivaluedness
  of the angle. For this, we display the fluctuations of the ensuing angle around the value $\pi/2$ as follows:
  in panel (a) we have details of  Figure \ref{fig-18}(a) over the frames 1500-2500, in panel (b) we have details of  Figure \ref{fig-18}(b) 
  over the frames 2000-3000, and in panel (c) we have details of Figure \ref{fig-18}(c) over the frames 500-1500.  }
  \la{fig-19}
\end{minipage}}
\end{figure}
 In each of the panels in Figure \ref{fig-19} we display the data in Figure \ref{fig-18}, over a subset of frames and now
 evaluated around the value $\pi/2$: in Figure \ref{fig-19}(a) we show the segment 1500-2500 
 of Figure   \ref{fig-18}(a), in Figure  \ref{fig-19}(b)
we show the segment 2000-3000 of Figure  \ref{fig-18}(b) and in Figure  \ref{fig-19}(c) we show the segment  500-1500
of Figure \ref{fig-18}(c).

By combining Figures \ref{fig-18} and \ref{fig-19} we conclude that there are  two major transitions, around frames 1000 and 2500
respectively. 
These transitions are concurrent with the major transitions in figures \ref{fig-8}, \ref{fig-15}, \ref{fig-17} and in particular
figure \ref{fig-9}.  The first transition corresponds to the creation of the N-terminal soliton, and the second one to its 
translation,  by one site towards the N-terminal, and subsequent stabilisation. We also observe the presence 
of an extended transition process,  visible in Figure \ref{fig-18}(b) between frames 1100-1500. In summary, we 
conclude from these figures that:

$\bullet$ In the vicinity of the frame 1000, when the N-terminal soliton structure forms, 
there is an initial twisting of the $i=5$ dihedral which is close to $+\pi$ and a twisting   
of the $i=6$ dihedral by an approximatively equal amount but in the opposite direction. Thus, at this point 
the total twisting which is produced along the side chain
segment vanishes. We note that the presence of two twists by an equal amount but opposite in direction, is
the hallmark of a Bloch domain wall  pair production.  But we recall that the backbone folding index detects only
a single soliton, as shown in Figure \ref{fig-9}(a). 

 $\bullet$ After the initial Bloch wall pair formation, the $i=5$ dihedral becomes slowly 
 twisted back to the original value between frames 1200-1600, so that after frame 1600 we are 
 left with a total of $\sim -\pi$ twist. This process leaves us with a  
 single soliton along the chain segment, with a total twisting around $-\pi$. 

$\bullet$ Finally, in the vicinity of frames  2100-2200, we observe a rapid twisting of the $i=2$ dihedral by an amount close to
$\approx -\pi$. This is followed, 
in the vicinity of the frame 2400-2500,  by a twisting of the $i=5$ dihedral by an approximatively 
equal amount. There is an accompanying twist of the $i=6$ dihedral by an amount somewhat less than $-\pi/2$, over the same frames. These two  twistings at $i=5,6$ accompany the Peierls-Nabarro barrier crossing, as can be deduced by comparison
with figures \ref{fig-15} and \ref{fig-17}.

$\bullet$ When the soliton stabilises,  after frame 3000, we conclude that there is a total twisting in the side chain
structure which is close to $\approx -3 \pi$,  and carried by the final soliton configuration.  Thus we assign to the final
soliton the value $Ind_m = -3$ of the index (\ref{eq:etaprime2}).

Note that there is also certain (small) spillage of the $\eta$-values which is distributed among the nearby residues.

In summary, the side chain analysis shows that at the level of the XY spin chain analysis, 
the  N-terminal soliton structure forms by an initial rapid formation of a Bloch domain wall {\it i.e.} 
soliton-antisoliton pair, followed by a slow twisting that apparently removes one of the two Bloch wall solitons.
This is then followed by a rapid transition, in combination
with a Peierls-Nabarro barrier crossing, that forms the  final stable soliton structure.
Accordingly, we may characterise the final soliton as a configuration with  
the C$\alpha$ backbone folding index $Ind_f = -1$ and the side chain XY folding index $Ind_m = -3$.

\subsubsection{Backbone}

Figure \ref{fig-13} revels the presence of strong correlations between backbone and side chain dynamics.
In particular,  any formation of side chain soliton structure should correlate with the formation of corresponding
backbone soliton. Accordingly, we proceed to construct explicitly the backbone soliton that 
accompanies the N-terminal side chain soliton. We  shall find that the backbone soliton can be 
modeled by a solution of the discrete nonlinear Schr\"odinger equation  (\ref{tauk}), (\ref{nlse}), 
with very high sub-atomic precision.
%Accordingly, the energy function (\ref{E1old}) does not only describe the stationary, 
%crystallographic minimum energy protein structures. It is
%also applicable for modelling the intermediate structures, that are found when a protein folds. 

As an example, we  consider the profile of the N-terminal soliton structure at two different frames. 
Other frames, and the C-terminal soliton structure,  can be  analysed similarly. We select the sites 4-11
(PDB sites 632-639) for our analysis: The sites 0-3 are subject to fluctuations, 
and beyond the site 11 there is only a monotonic $\alpha$-helix. 

We start with the frame 2000, which is located in the  regime where the side chain structure   
of the soliton has stabilised according to Figure \ref{fig-18}(b) and the backbone folding index of the segment 
shown in fFigure \ref{fig-9}(a) has the value  $Ind_f = +1$; the side chain index over this segment is $Ind_m = -1$,
according to Figures \ref{fig-18}, \ref{fig-19}.

In Figure \ref{fig-20} we show the profile of the bond angle and the torsion angle over the segment 4-11 (sites 632-639 in the PDB file) in the frame 2000,
after we have implemented the $\mathbb Z_2$ gauge transformation (\ref{eq:dsgau}) to identify the
soliton profile. Note that in order to compute a single bond angle, we need to know three residues while the evaluation of a torsion
angle consumes four residues. Thus, despite the smaller number of data points in the figure, the ensuing configuration engages
8 residues.
%
%
%                                                                         FIGURE 20
%
%
\begin{figure}
\centering
{\begin{minipage}[h]{.45\textwidth}
  \raggedright
  \includegraphics[trim = 0mm 5mm 20mm -10mm, width=0.9\textwidth,  angle=0]{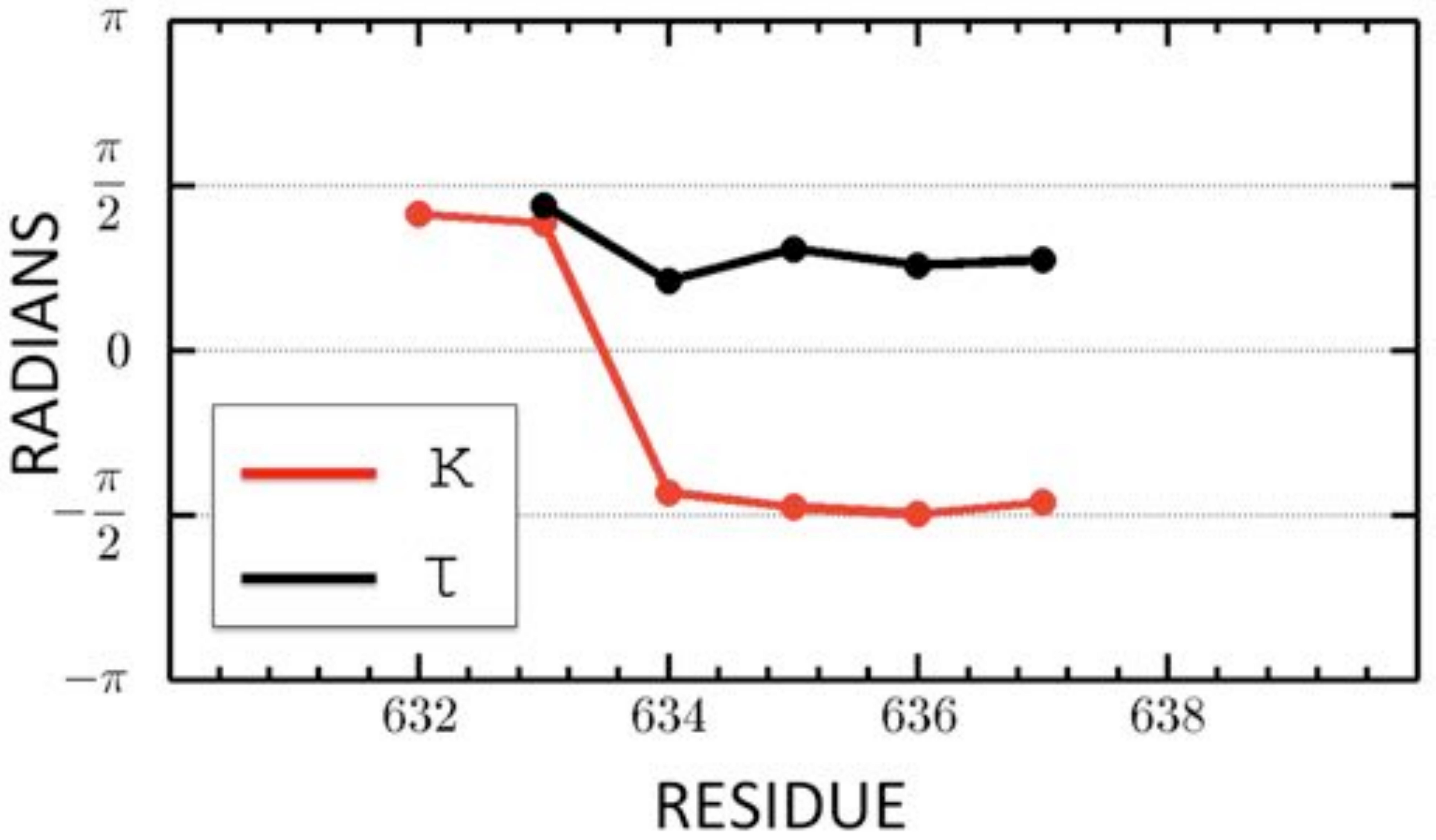}
  \caption{The $\mathbb Z_2$ gauge-transformed bond angles $\kappa_i$ (red)  and torsion angles (black) 
  for the segment 4-11 (PDB index 632-639) in frame 2000.  Note that bond angle takes 3 residues, and
  torsion angle takes 4 residues to compute.}
  \la{fig-20}
\end{minipage}}
\end{figure}

We observe that the bond angle has the profile of a single domain wall soliton of the DNLS equation, approximated by 
(\ref{kappaprof}).  We use the software package {\tt ProPro} that has been described
at 
\[
{\tt http://www.protein-folding.org}
\]
to numerically construct the ensuing soliton solution of the DNLS equation.

In Figure \ref{fig-21} (a) we compare the profile of the bond and torsion angles in Figure \ref{fig-20} with  
the profile of the soliton solution of  (\ref{tauk}), (\ref{nlse}).
%
%
%                                                                         FIGURE 21
%
%
\begin{figure}
\centering
{\begin{minipage}[h]{.45\textwidth}
  \raggedright
  \includegraphics[trim = 0mm 5mm 20mm -10mm, width=0.9\textwidth,  angle=0]{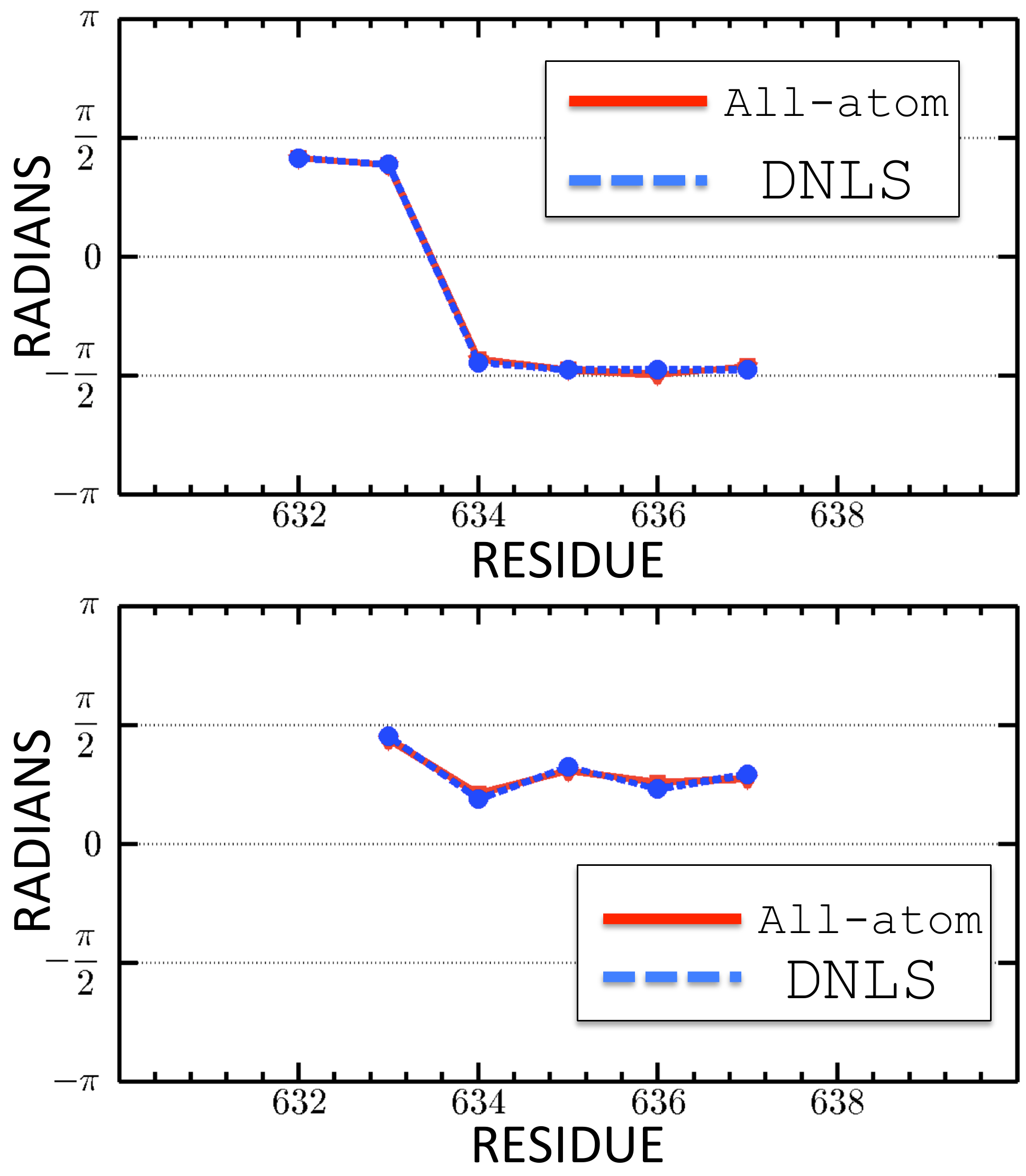}
  \caption{The $\mathbb Z_2$ gauge-transformed all-atom bond angles $\kappa_i$ (top)  and torsion angles (bottom) 
  for the segment 632-639 in frame 2000, compared with the corresponding DNLS soliton, Eqs. (\ref{tauk}), (\ref{nlse}). }
  \la{fig-21}
\end{minipage}}
\end{figure}

In Figure \ref{fig-22} (top) we compare the residue-wise distance between the all-atom configuration of frame 2000,
and the DNLS solution.
The average C$\alpha$ RMSD 
between the two configurations is less than 0.1 \AA ngstr\"om, and at no residue is the distance between the C$\alpha$-atoms more
than 0.2  \AA ngstr\"om: the difference is truly negligible. The
grey strip around the DNLS soliton is a 0.2 \AA ngstr\"om (quantum mechanical) fluctuation band \cite{Krokhotin-2011}. 
The Figure  \ref{fig-22} (bottom) shows the 3D overlay of the
ensuing C$\alpha$ backbones, for all practical purposes they are the same. 
%
%
%                                                                         FIGURE 22
%
%
\begin{figure}
\centering
{\begin{minipage}[h]{.45\textwidth}
  %\raggedright
  \includegraphics[trim = 0mm 5mm 20mm -10mm, width=0.9\textwidth,  angle=0]{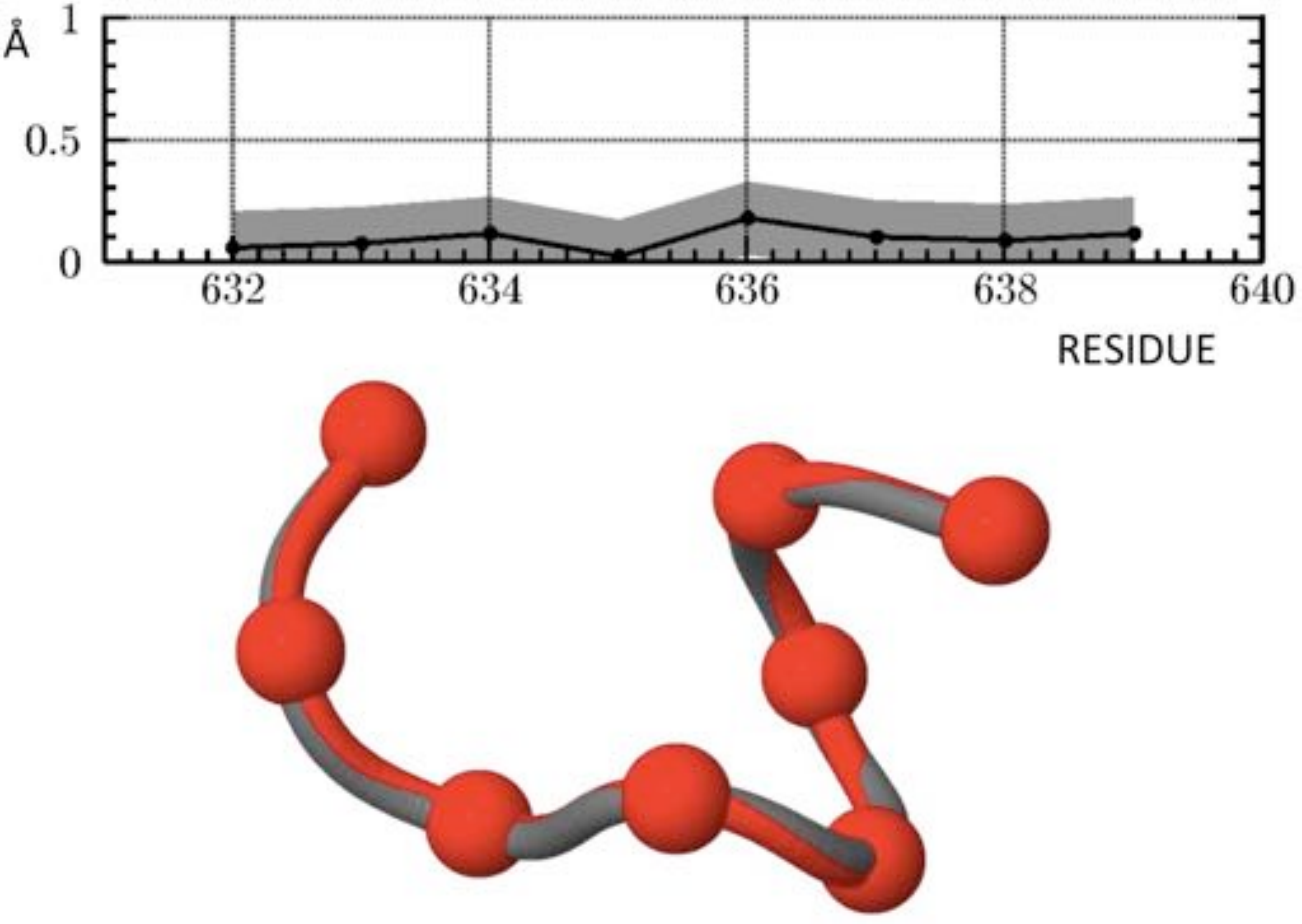}
  \caption{Top panel: The residue-wise distance between C$\alpha$ backbone of the MD simulated soliton at frame 2000, and
  the DNSL soliton solution; the grey strip represents the estimated 0.2 \AA~ quantum
  mechanical fluctuation band; Bottom panel: The 3D superimposition of the all-atom structure (grey) with the DNLS  soliton (in red).}
  \la{fig-22}
\end{minipage}}
\end{figure}

In Figure \ref{fig-23} we show the loop trajectories of the all-atom configuration of frame 2000, and the corresponding
DNLS soliton, on the stereographically projected two-sphere of Figure \ref{fig-2} and \ref{fig-3}. 
%
%
%                                                                         FIGURE 23
%
%
\begin{figure}
\centering
{\begin{minipage}[h]{.45\textwidth}
  %\raggedright
  \includegraphics[trim = 0mm 5mm 20mm -10mm, width=0.9\textwidth,  angle=0]{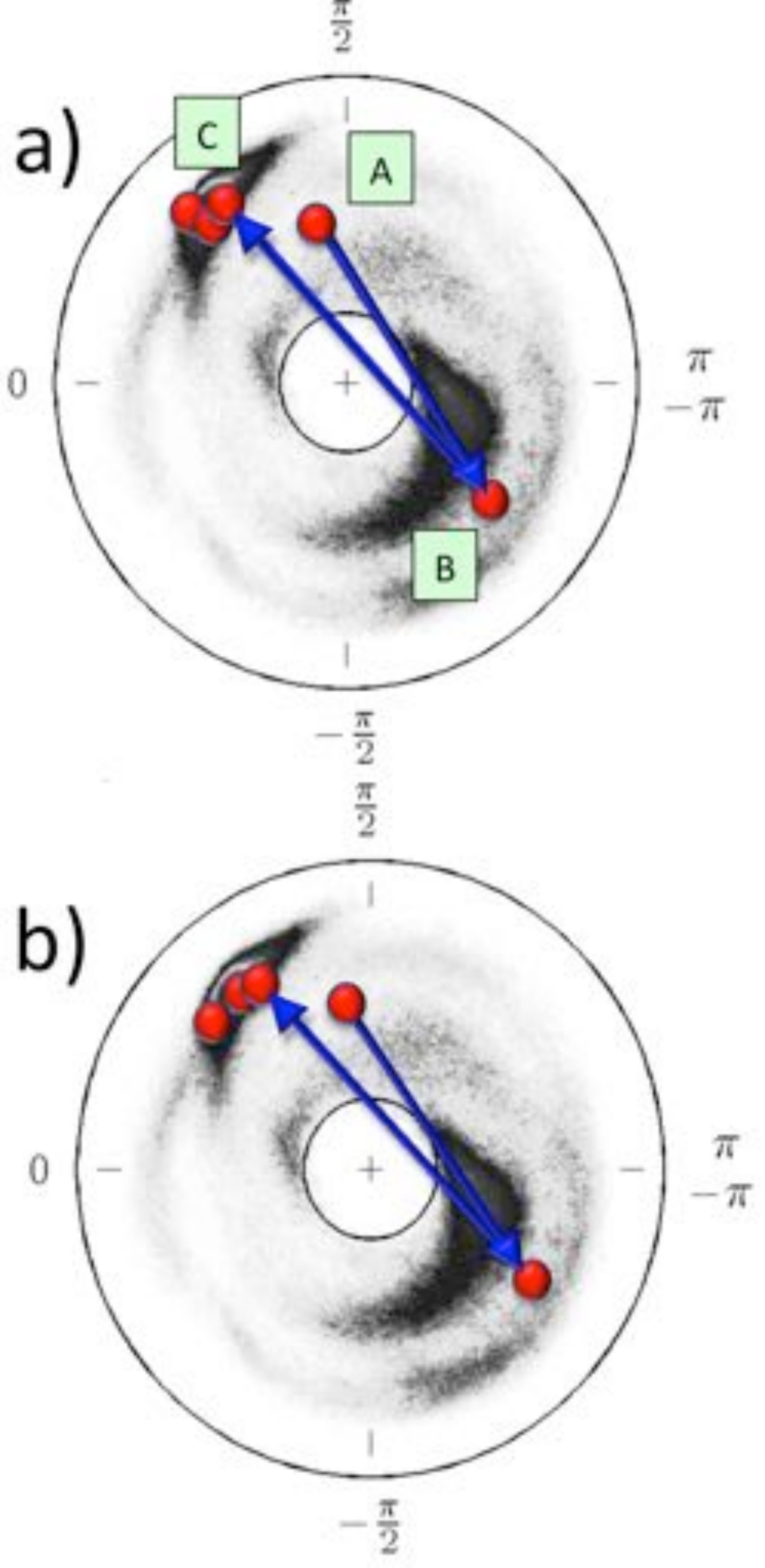}
  \caption{top: The loop trajectories  of Figure \ref{fig-3} (a) in the case of: (a) the all-atom frame 2000 segment 632-639, and (b) the corresponding DNLS soliton.}
  \la{fig-23}
\end{minipage}}
\end{figure}
Note that for both of these two  loop trajectories the folding index, as defined in (\ref{eq:Gamma}) is vanishing, in that the trajectory
does not encircle the north-pole (center of the disk). A very short change either in the position of the residue labeled $B$ or 
in the position of the residue labeled C in Figure \ref{fig-23} a), can shift the
trajectory so that the line connecting them moves over to the other side of the north pole and the folding index 
becomes $Ind_f = +2$. This is consistent with the result shown in Figure \ref{fig-9} that the folding
index fluctuates between the values $Ind_f = 0$ and $Ind_f = +2$, around the frame 2000, with the posture shown in Figure
\ref{fig-23} being the more stable one. 

In Figure \ref{fig-24} we show a close-up to the frame segment 2475-2525 around the soliton 2000, in terms of the side chain angles
$\eta$. The  close-up reveals the presence of fluctuations between the soliton and the N-terminal, while the helix between the soliton
and the C-terminal displays very small fluctuations. Thus, the fluctuations in the folding index around the frame 2000 are most likely due
to shifts in the position of the residue labeled B  in Figure \ref{fig-23} (a).
%
%
%                                                                         FIGURE 24
%
%
\begin{figure}
\centering
{\begin{minipage}[h]{.45\textwidth}
  \raggedright
  \includegraphics[trim = 0mm 5mm 20mm -10mm, width=0.9\textwidth,  angle=0]{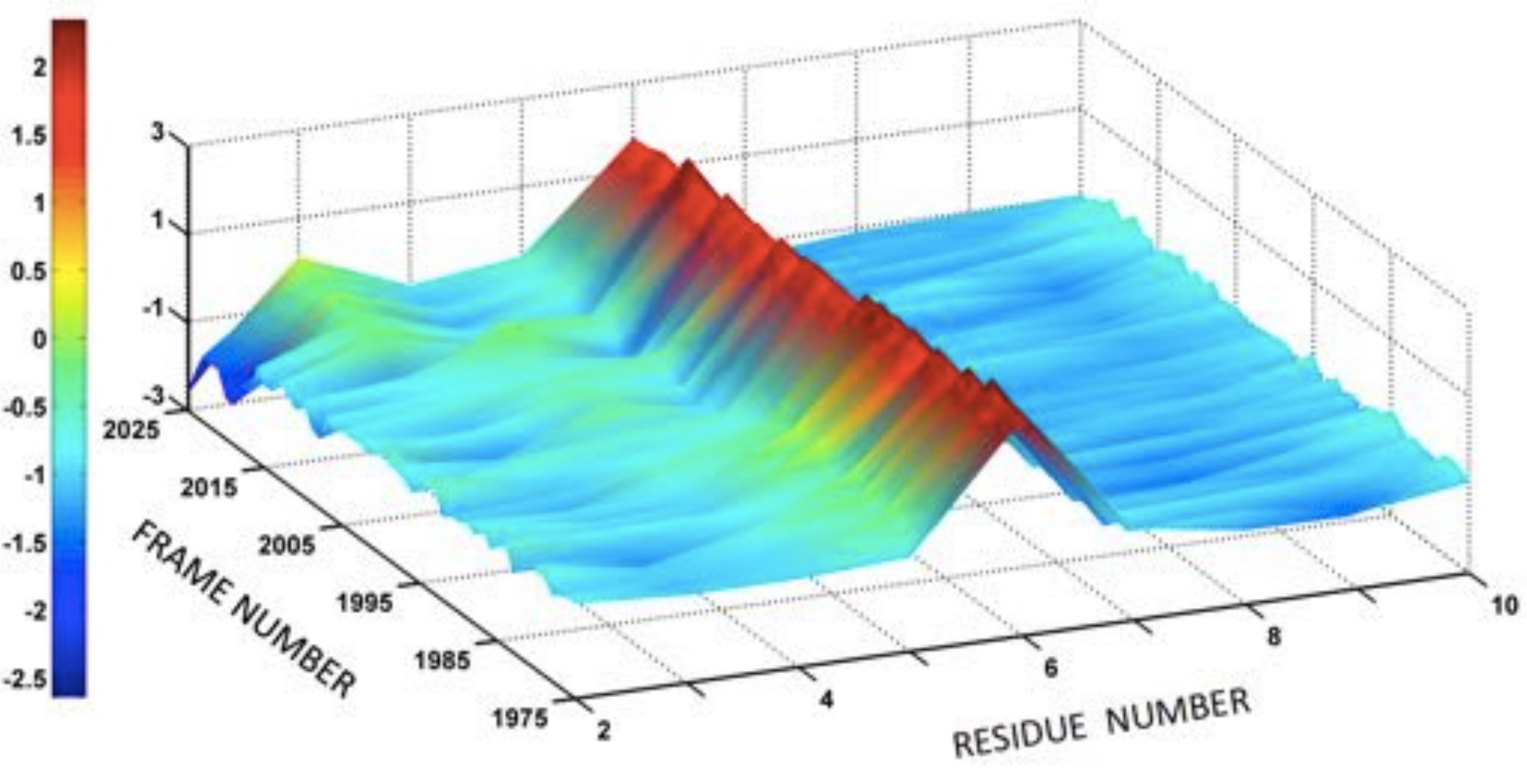}
  \caption{A close-up of Figure \ref{fig-17}, around the frame 2000. The $\alpha$-helical structure between the soliton and the C-terminal displays only very slight fluctuations, while the fluctuations between  the soliton and the N-terminal are more profound.}
  \la{fig-24}
\end{minipage}}
\end{figure}

Figures \ref{fig-25}-\ref{fig-28} show the same analyses for the conformation in frame 3500.
%
%
%
%
%                                                                         FIGURE 25
%
%
\begin{figure}
\centering
{\begin{minipage}[h]{.45\textwidth}
  \raggedright
  \includegraphics[trim = 0mm 5mm 20mm -10mm, width=0.9\textwidth,  angle=0]{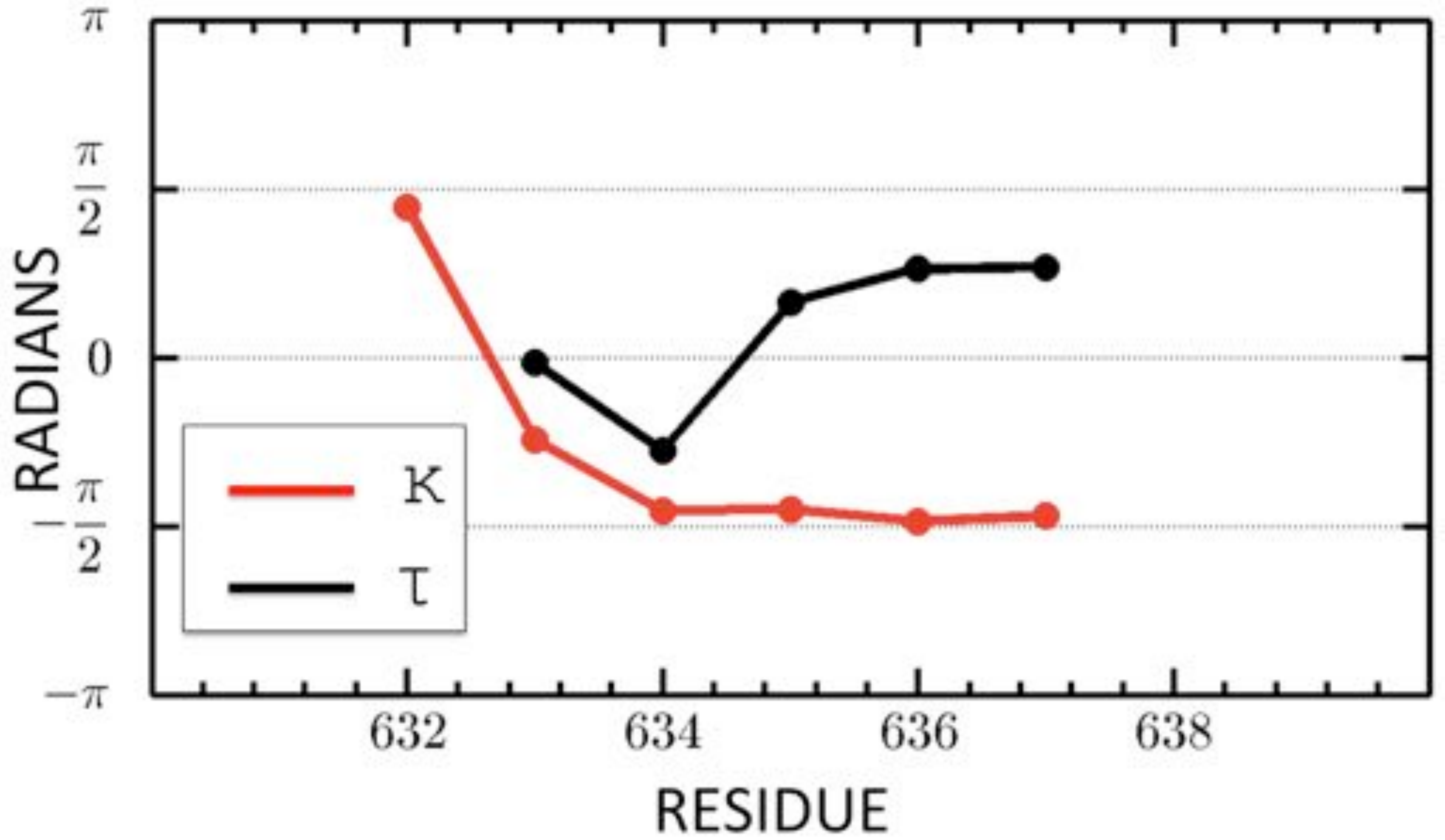}
  \caption{Same as in Figure \ref{fig-20}, for the frame 3500.}
  \la{fig-25}
\end{minipage}}
\end{figure}
%
%
% 
%
%
%                                                                         FIGURE 26
%
%
\begin{figure}
\centering
{\begin{minipage}[h]{.45\textwidth}
  \raggedright
  \includegraphics[trim = 0mm 5mm 20mm -10mm, width=0.9\textwidth,  angle=0]{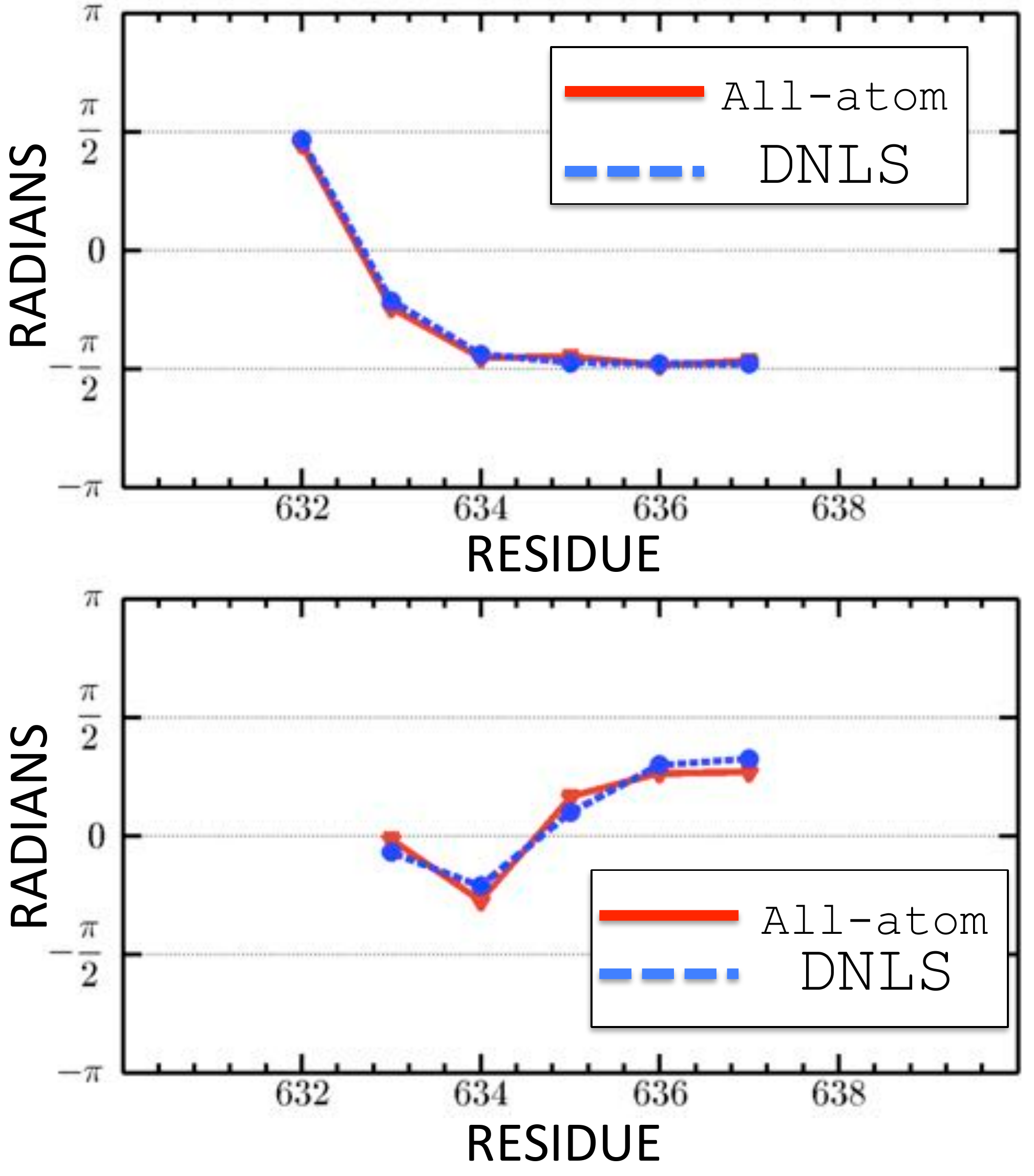}
  \caption{Same as in Figure \ref{fig-21}, for the frame 3500.}
  \la{fig-26}
\end{minipage}}
\end{figure}
%
%
% 
%
%
%                                                                         FIGURE 27
%
%
\begin{figure}
\centering
{\begin{minipage}[h]{.45\textwidth}
  \raggedright
  \includegraphics[trim = 0mm 5mm 20mm -10mm, width=0.9\textwidth,  angle=0]{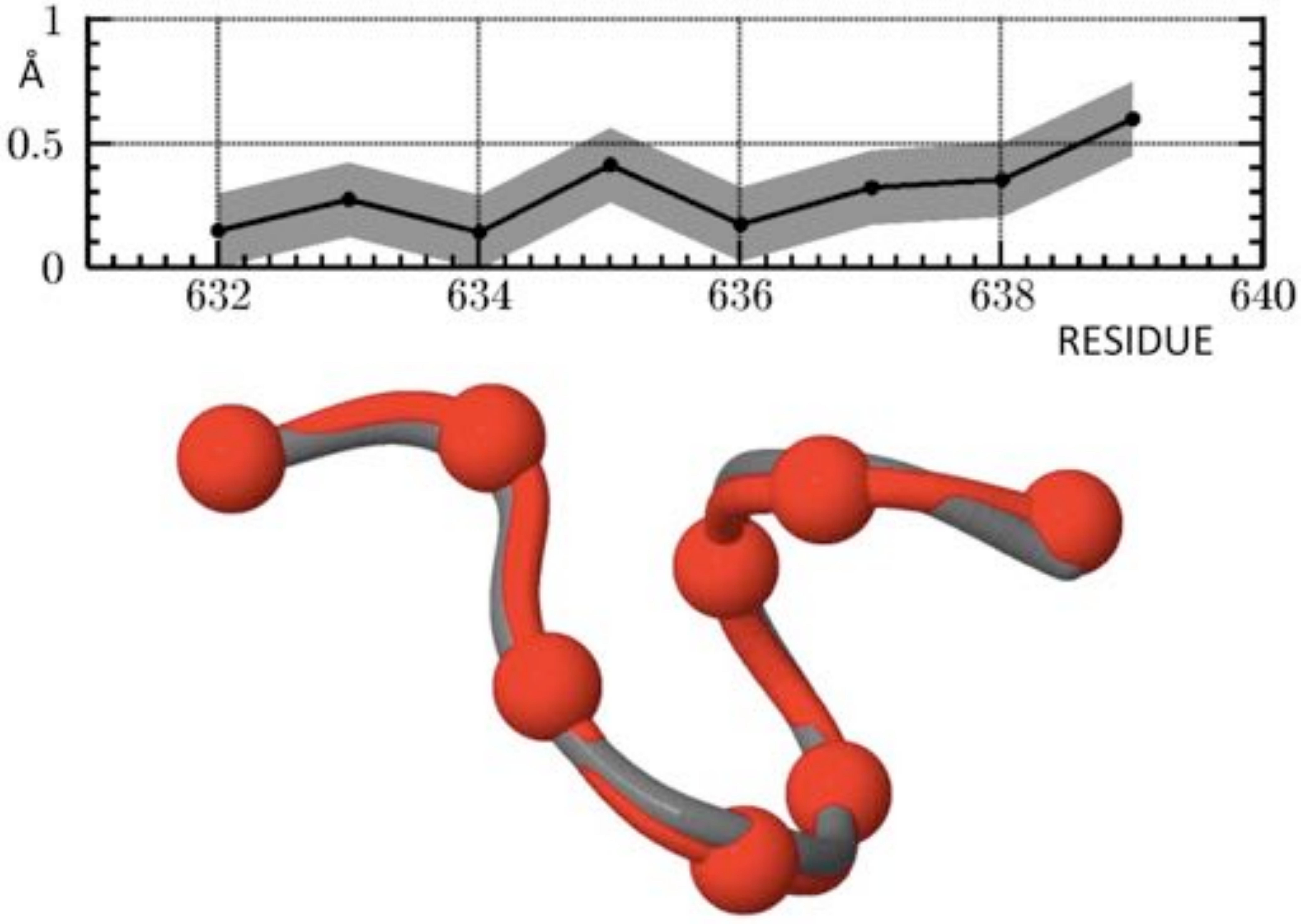}
  \caption{Same as in Figure \ref{fig-22}, for the frame 3500.}
  \la{fig-27}
\end{minipage}}
\end{figure}
%
%
% 
%
%
% 
%
%
%                                                                         FIGURE 28
%
%
\begin{figure}
\centering
{\begin{minipage}[h]{.45\textwidth}
  %\raggedright
  \includegraphics[trim = 0mm 5mm 20mm -10mm, width=0.9\textwidth,  angle=0]{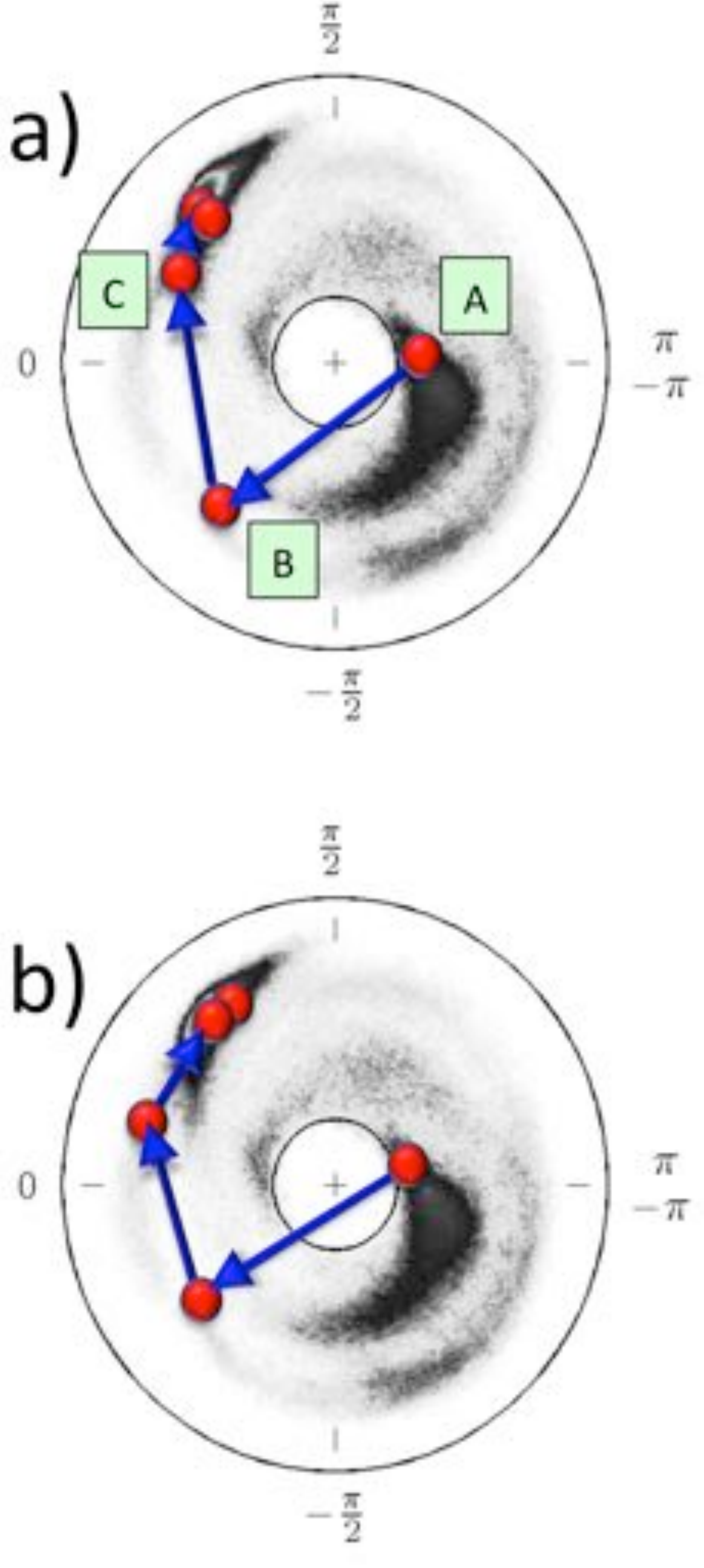}
  \caption{Same as in Figure \ref{fig-23}, for the frame 3500.}
  \la{fig-28}
\end{minipage}}
\end{figure}
We find that the DNLS soliton describes the domain wall soliton
that  we have constructed by all-atom simulations, with a very high sub-atomic
precision. We note that the soliton in frame 3500 
is a configuration that connects between the $\beta$-stranded region of Figure \ref{fig-2}
to the $\alpha$-helical region, while in the case of the soliton at frame 2500 the initial residue is located in a sparsely populated region
of the landscape in Figure \ref{fig-2}. 
% We also note that the trajectory in frame 3500 is very similar to the steps 5, 6 and 7 in the frame 4000 trajectory, shown in Figure \ref{fig-12}; the steps 1-4 in  \ref{fig-12} correspond to the N-terminal residues.
 
The major qualitative difference between frames 2000 and 3500 
is between Figures \ref{fig-23} and \ref{fig-28}. The soliton in frame 3500 is relatively
stable. In particular, as shown in Figure \ref{fig-9}, its folding index $Ind_f = +1$. We can understand the stability of the folding index
by comparing Figure \ref{fig-23} (a) with Figure \ref{fig-28} (a). In the later, the residues have assumed positions where the connecting
arrows are stabilised against small perturbations, they are protected from  crossing over the north pole in a manner that causes
fluctuations in the value of the folding index.

Finally, we compare   the structure of the solitons at frames 2500 and 3500. From Figures \ref{fig-20}, \ref{fig-25} we observe
that in terms of the bond angles the soliton 3500 has indeed moved one site towards the N-terminal, from the position of 
soliton 2500.  In Figure \ref{fig-29} we overlay their 3D structures. For this, we first translate the soliton in frame 3500
one residue away from the N-terminal, so that the two have the same location along the backbone. The figure shows the 
ensuing 3D interlaced C$\alpha$ backbones, in a relative position where the RMSD is minimal. 
There is a visible difference, and the minimal RMSD is 2.0 \AA; the soliton has clearly become deformed. 
%
% 
%
%
%                                                                         FIGURE 29
%
%
\begin{figure}
\centering
{\begin{minipage}[h]{.42\textwidth}
  %\raggedright
  \includegraphics[trim = 0mm 5mm 20mm -10mm, width=0.9\textwidth,  angle=0]{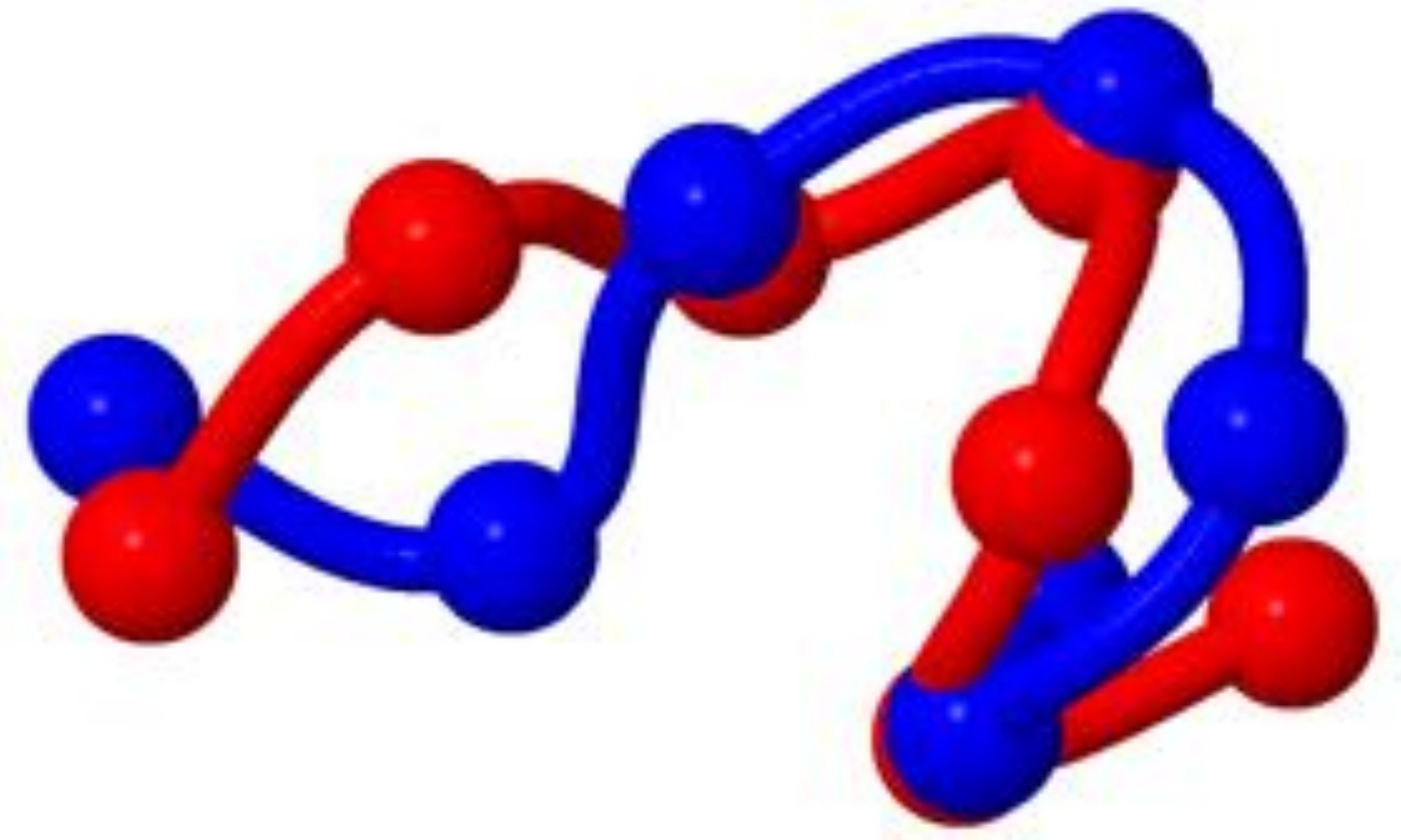}
  \caption{Comparison of the solitons at frame 2500 (red) and 3500 (blue), after the second has been translated back to the 
  location of the first one. The RMDS is 2.0 \AA.}
  \la{fig-29}
\end{minipage}}
\end{figure}

\section{Summary}

Molecular dynamics enables the scrutiny of protein folding, at the level of individual atoms and 
over very short time intervals.  However, it can  leave us with the conceptual challenge to understand, 
how the individual atoms cooperate to produce the kind of large scale organisation that appears to be
prevalent among crystallographic protein structures.

We have performed detailed molecular 
dynamics simulations, with the aim to find out how organised structure emerges when a protein folds.  
We have first compared three different force fields using the GROMACS 4.6.3. 
package, to select the proper tools.  We have chosen a C-chain subunit from HIV envelope glycoprotein with PDB code 
1AIK as a concrete example, partly due to its biomedical relevance even though this is an  issue which has
not been addressed by us. We have  introduced and further developed various tools of modern
theoretical physics, to systematise and analyse the data. These include topological tools, conceptual analogies drawn from the
notion of spin chains,  the notion of Wilsonian universality, and methods based on the analytical 
structure of the discrete nonlinear  Schr\"odinger equation. In this manner we have arrived at the  conclusion 
that the protein folding is a process that relates intimately to the emergence and interactions of solitons. In particular, a configuration such as the Bloch domain wall along a spin chain
appears to be most useful in comprehending how structure emerges and self-organises when a protein folds. 

We have inspected both the
static and dynamic properties of domain wall solitons and observed that concepts which are familiar from the study of
lattice systems, such as the Peierls-Nabarro barrier, also appear along protein backbones lattices, 
and in fact assume a central role
in dictating how the folding proceeds. We hope that our observations 
help to pave a way for the powerful analytical and
topological tools and techniques that have been introduced and developed in the context of integrable spin chains and related
solvable models, to become part of the arsenal used  describe emergence of structure and organisation in the case of proteins
and other biological macromolecules.

\section{Acknowledgements:}
This research was supported in part by Bulgarian Science Fund (Grant DNTS-CH 01/9/2014) and 
China-Bulgaria Intergovernmental S\&T Cooperation Project at Ministry of Science and Technology of P.R. 
China (2014-3). AJN acknowledges support from Vetenskapsr\aa det, Carl Trygger's Stiftelse f\"or vetenskaplig forskning 
and  Qian Ren Grant at BIT, P.R. China. AS was supported by National Science Center Poland (Maestro UMO-2012/06/A/ST4/00376).

\end{document}